\documentclass{aa}
\usepackage{newtxtext,newtxmath}
\usepackage[T1]{fontenc}
\usepackage{graphicx}
\usepackage{amsmath}
\usepackage[normalem]{ulem}
\usepackage{orcidlink}
\usepackage{enumitem}
\usepackage{pdflscape}
\usepackage{siunitx}

\usepackage{lipsum}
\usepackage{subcaption}

\DeclareMathOperator{\logten}{log_{10}}

\newcommand{\Msun}{M_{\sun}}
\newcommand{\Mhalo}{M_{\mathrm{200}}}
\newcommand{\logMhalo}{\logten(\Mhalo/\Msun)}
\newcommand{\disturbedvotes}{$\mathrm{V}_\mathrm{dist.}$}
\newcommand{\disturbedfrac}{$F_\mathrm{dist.}$}
\newcommand{\undisturbedfrac}{$F_\mathrm{undist.}$}
\newcommand{\undisturbedvotes}{$\mathrm{V}_\mathrm{undist.}$}
\newcommand{\problemvotes}{$\mathrm{V}_\mathrm{prob.}$}
\newcommand{\problemfrac}{$F_\mathrm{prob.}$}
\newcommand{\totalvotes}{$\mathrm{V}_\mathrm{total}$}
\newcommand{\mergervotes}{$\mathrm{V}_\mathrm{merg.}$}
\newcommand{\mergerfrac}{$F_\mathrm{merg.}$}

\newcommand{\tailvotes}{$\mathrm{V}_\mathrm{tail}$}
\newcommand{\tailfrac}{$F_\mathrm{tail}$}
\newcommand{\disturbedfractioneqn}{\disturbedvotes / (\disturbedvotes + \undisturbedvotes)}
\newcommand{\undisturbedfractioneqn}{\undisturbedvotes / (\disturbedvotes + \undisturbedvotes)}

\newcolumntype{F}{S[round-mode=places,round-precision=2]}
\newcolumntype{V}{S[round-mode=places,round-precision=0,table-format=2]}
\newcolumntype{W}{S[round-mode=places,round-precision=1]}

\begin{document}


\title{Fishing for Jellyfish Galaxies:\\Exploring ram-pressure stripping with crowd science}

   \author{C. Bellhouse\inst{1,2}
        \and Y.~L. Jaff\'e\inst{3,4}\fnmsep\thanks{Corresponding author: yara.jaffe@usm.cl}
        \and J. P. Crossett\inst{3}
        \and S. L. McGee\inst{5}
        \and Ignacio Quiroz\inst{3}
        \and C. Dulcien\inst{3}
        \and R. Smith\inst{3}
        \and B. M. Poggianti\inst{1}
        \and B. Vulcani\inst{1}
        \and V. Sampaio\inst{3}
        \and A. Werle\inst{1}
        \and N. Tomi\^ci\'{c}\inst{6}
        \and A. Müller\inst{7}
        \and N. Akerman\inst{1,8}
        \and A. Ignesti\inst{1}
        \and A. Khoram\inst{9,10}
        }

\institute{
INAF- Osservatorio astronomico di Padova, Vicolo Osservatorio 5, I-35122 Padova, Italy
\and School of Physics and Astronomy, University of Nottingham, University Park, Nottingham NG7 2RD, UK
\and Departamento de Física, Universidad Técnica Federico Santa María, Avenida España 1680, Valparaíso, Chile
\and Millennium Nucleus for Galaxies (MINGAL)
\and School of Physics and Astronomy, University of Birmingham, Birmingham B15 2TT, UK
\and Department of Physics, Faculty of Science, University of Zagreb, Bijenička Cesta 32, 10000 Zagreb, Croatia
\and Ruhr University Bochum, Faculty of Physics and Astronomy, Astronomical Institute (AIRUB), Universitätsstraße 150, 44801 Bochum, Germany
\and Dipartimento di Fisica e Astronomia ‘Galileo Galilei’, Università di Padova, vicolo dell’Osservatorio 3, IT-35122 Padova, Italy
\and Dipartimento di Fisica e Astronomia, Universit`a di Bologna, Via Gobetti 93/2, I-40129, Bologna, Italy
\and INAF, Astrophysics and Space Science Observatory Bologna, Via P. Gobetti 93/3, I-40129 Bologna, Italy
}

\date{Received September 30, 20XX}

  \abstract
   {}
   {We present the first results of \textit{Fishing for Jellyfish Galaxies}, a pilot citizen-science project using Zooniverse to identify galaxies undergoing ram-pressure stripping (RPS).}
   {Volunteers visually inspected colour images of late-type galaxies from the Dark Energy Camera Legacy Survey. The sample consisted of 49,703 galaxies selected within $4 \times R_{500}$ of clusters and groups, restricted to those brighter than 19th magnitude in the \textit{g} and \textit{r} bands, and with a minimum half-light radius of 2 arc seconds, to aid classification. We detail our data processing, including debiasing classifications and optimising vote-fraction thresholds to maximise completeness and purity, calibrated against a ground-truth set of pre-labelled galaxies.}
   {Our final catalogue contains 6739 jellyfish candidates (6621 new), 5430 merger candidates, and 29,729 undisturbed galaxies, with 3910 jellyfish exhibiting prominent tail-like morphologies. We find that the fraction of RPS candidates rises from $\sim$10\% in galaxy groups to $\sim$20--30\% in massive clusters, confirming the findings of previous studies carried out on smaller samples. For the subset of our RPS candidate sample with spectroscopic data, we measure a median cluster-centric velocity $53\%$ higher than the general cluster population, consistent with galaxies in early stages of accretion into the cluster. They are also typically late-type blue galaxies with elevated star-formation rates, in agreement with expectations. These results demonstrate that citizen scientists can reliably identify galaxies that undergo environmental processes. We provide the initial release of 37,599 visually classified galaxies as a resource for future studies of galaxy transformation in clusters.}
   {}


\maketitle
\nolinenumbers
\section{Introduction}

Dense environments, such as those found in groups and clusters, can promote gravitational and hydrodynamical encounters which drastically alter the morphology and gas content of a galaxy \citep{Boselli2006}. Since gas fuels star formation, understanding these processes and their influence on the population of a cluster is critical to the study of galaxy evolution.

The mechanisms which transform galaxies in clusters can be divided into two primary categories: gravitational effects, resulting from tidal encounters between galaxies \citep{Spitzer1951,Toomre1977,Tinsley1979,Merritt1983,Mihos1993,Springel2000} or their interactions with the cluster potential \citep{Byrd1990,Valluri1993}, and hydrodynamical processes, resulting from the interaction between galaxies and the intracluster medium (ICM).

One of the most efficient mechanisms of gas removal is the process of ram-pressure stripping \citep[RPS;][]{Gunn1972}. When a galaxy infalls into a cluster with sufficient velocity, the cluster ICM imparts a drag force that acts upon the interstellar medium within the galaxy. The resulting interaction can effectively remove the gas component from the galaxy, forming a trail of material behind it, without directly disrupting the existing stellar component, which is too dense to be affected. The observable effects of RPS include leading-edge compression \citep{Vollmer2001}, the presence of a tail or trailing material, the condensation of star-forming clumps in the tail \citep{Kenney2004,Fossati2016, Giunchi2023a,Giunchi2023b,Giunchi2025} accompanied by a temporary boost in star-formation \citep{Poggianti2016,Vulcani2018,Vulcani2020,Vulcani2023, Tomicic2018}, eventually leading to quenching if the gas is completely removed. In such cases, galaxies typically transition into early-type morphologies \citep{Sampaio2022,Sampaio2024}. In addition, several cases of galaxies being `unwound' by RPS have been observed \citep{Bellhouse2021,Vulcani2022,Matijevic2026}.

The most extreme cases of RPS result in so-called `jellyfish' galaxies, with characteristic tails of stripped material. Such galaxies are valuable laboratories for testing the processes that enhance and quench star formation in galaxies \citep{Poggianti2019b,Vulcani2022} and, in rare cases, may trigger active galactic nucleus (AGN) formation \citep{Poggianti2017b, Peluso2022}.
Untangling the complex relationship between hydrodynamical influences, morphology, and environment requires a large sample of galaxies. This is especially important when measuring the effects of environment in subsamples of galaxies with specific properties, which may constitute only a small fraction of the full sample.

Statistical studies of RPS across cluster properties such as halo mass and dynamical state have highlighted the need for larger, homogeneous samples of stripped galaxies. Ram-pressure stripping has been shown to be enhanced by cluster mergers and shocks in simulations \citep{Roediger2014} and individual clusters \citep{Stroe2015,Stroe2020,Bellhouse2022}. \citet{Lourenco2023} extended this analysis across a larger sample of clusters, finding an increase in instances of RPS with increased cluster disturbance; however, they emphasise that larger homogeneous samples of RPS galaxies are necessary to provide a statistically significant result. Similarly, \citet{Salinas2024} carried out statistical studies of the tail directions in RPS galaxies and highlighted the importance of increased sample sizes, with a particular emphasis on homogeneous samples of spectroscopically confirmed cluster members.

A major challenge in the study of RPS galaxies is identification. Past samples of RPS galaxies have been identified in various ways,
from visual inspection of imaging data in optical \citep{Poggianti2016,McPartland2016,Roberts2020,Vulcani2022,Kolcu2022} and UV \citep{George2024} wavelengths, H$\alpha$ imaging \citep{Yagi2010}, HI and radio continuum surveys \citep{Roberts2022, Ignesti2023}, or X-ray data \citep{Sun2006}. A discussion of the different selection criteria and their inherent drawbacks and advantages can be found in \citet{Poggianti2025}. The disturbance signatures are generally morphological and widely varied in their nature, making automated classification a challenge, especially given the relatively small current sample. One practical solution to this lies in citizen science projects. The success of Galaxy Zoo \citep{Lintott2008, Walmsley2022, Bamford2009, Willett2013} and its Zooniverse toolkit has given rise to dedicated citizen science projects. One such example is \citet{Zinger2024}, which showed images of simulated galaxies from IllustrisTNG \citep{Vogelsberger2014, Genel2014, Sijacki2015} to thousands of volunteers, who were asked to classify whether their visual appearance was consistent with galaxies experiencing RPS. Until now, such an analysis has not been carried out on observational data. By combining the responses of several volunteers for each galaxy, we can separate different galaxy types based on vote fractions. \citet{Crossett2025} showed that even when the volunteers are not specifically looking for signs of RPS, their combined responses on questions about `oddities' can be used to identify the presence of RPS signatures in images of galaxies.

Building on the success of these volunteer-powered projects and with the goal of producing a large, homogeneous sample of visually classified stripping candidates (SCs) for future statistical studies of RPS, we designed and executed a dedicated citizen science project based on the Zooniverse platform. In Sect.\,\ref{sec:sample} we detail the process of selecting target clusters across a range of masses and the selection of satellite galaxies therein, outlining the morphology, brightness, and angular size constraints on the subject galaxies. Section\,\ref{sec:classifications} describes the classification process, including details of the Zooniverse workflow, definitions of the vote fractions, as well as the debiasing and combination of classifications to produce the final sample selection criteria. Section\,\ref{sec:catalogue} describes the catalogue, detailing the included parameters and showing examples of jellyfish candidate galaxies found by the identification process.
To test the viability of the visually classified sample of galaxies as genuine candidates for stripping as well as investigating trends in the incidence of RPS across cluster mass, we carried out an initial analysis of the results, described in Sect.\,\ref{sec:results}. We investigate the abundance of RPS candidates as a function of environment and compare the morphologies and colours of the different samples and the star formation rates (SFRs) of a subset of the sample cross-matched with SDSS. Finally, in Sect.\,\ref{sec:conclusions}, we summarise the results and draw conclusions. Throughout this paper we assume a standard Lambda cold dark matter ($\Lambda$CDM) flat cosmology with $\Omega_\mathrm{m}=0.3$ and $H_0=70\,\mathrm{km}\,\mathrm{s}^{-1}\,\mathrm{Mpc}^{-1}$.

\section{The sample}\label{sec:sample}

\begin{table}
    \centering
    \caption{Host clusters included in the DECaLS data release 9 sample.}
    \begin{tabular}{lllr}
    \hline
    Cluster & Redshift & $\mathrm{Log}_{10}(\mathrm{M}_{200})$ & N$_{\rm galaxies}$\\
    \hline
    Abell 2052 & 0.035 & 14.72 & 743 \\
    Abell 957* & 0.044 & 14.36 & 226 \\
    Abell 3301* & 0.054 & 14.40 & 177 \\
    RXJ0058.9+2657 & 0.045 & 14.23 & 141 \\
    Abell 3667* & 0.056 & 15.01 & 458 \\
    Abell 2877* & 0.024 & 14.16 & 637 \\
    Abell 3341* & 0.038 & 14.34 & 270 \\
    RXJ1740.5+3539 & 0.043 & 14.27 & 202 \\
    ZwCl8338 & 0.050 & 14.42 & 162 \\
    Abell 2151a & 0.037 & 14.41 & 404 \\
    Abell 2147 & 0.035 & 14.73 & 896 \\
    MKW 8 & 0.026 & 14.22 & 576 \\
    Abell 2572a & 0.042 & 14.53 & 282 \\
    Abell 576 & 0.038 & 14.56 & 549 \\
    Abell 2199 & 0.030 & 14.84 & 1106 \\
    Abell 76 & 0.040 & 14.43 & 248 \\
    Abell S560* & 0.037 & 13.98 & 108 \\
    Abell 496* & 0.033 & 14.82 & 436 \\
    Abell 2634* & 0.031 & 14.37 & 708 \\
    Abell 85* & 0.056 & 15.20 & 418 \\
    Abell 2063 & 0.035 & 14.63 & 576 \\
    CGCG170-018 & 0.050 & 14.59 & 83 \\
    MKW4 & 0.020 & 14.20 & 728 \\
    Abell S463 & 0.040 & 14.37 & 252 \\
    Abell 3391 & 0.051 &  & 265 \\
    NGC1407 & 0.006 & 13.34 & 1204 \\
    Abell 548W* & 0.042 & 14.07 & 234 \\
    Pegasus II & 0.042 & 14.38 & 270 \\
    Abell 3376* & 0.047 & 14.58 & 230 \\
    Abell S861 & 0.051 & 14.47 & 370 \\
    Abell 671 & 0.050 & 14.36 & 186 \\
    UGC 04991 & 0.032 &  & 139 \\
    Abell S41 & 0.049 & 14.46 & 218 \\
    Abell 548E* & 0.042 & 14.56 & 326 \\
    NGC 4636 & 0.003 & 13.76 & 4833 \\
    Abell 1367 & 0.021 & 14.53 & 1633 \\
    \end{tabular}
    \tablefoot{Clusters marked with an asterisk are also in the CHANCES catalogue, which will provide additional spectroscopic coverage for future analysis. The N$_{\rm galaxies}$ column describes the number of galaxies included in the subject catalogue, selected from the $4\times\mathrm{R}_{500}$ region surrounding the cluster.}
    \label{tab:dr9_clusters}
\end{table}
\begin{table}
    \centering
    \caption{Host clusters included in the DECaLS data release 10 sample.}
    \begin{tabular}{lllr}
    \hline
    Cluster & Redshift & $\mathrm{Log}_{10}(\mathrm{M}_{200})$ & N$_{\rm galaxies}$\\
    \hline
    Abell 3562 & 0.049 & 14.71 & 460 \\
    Abell S727 & 0.050 & 14.46 & 108 \\
    Abell 3581 & 0.023 & 14.37 & 616 \\
    Abell S753 & 0.013 & 13.89 & 682 \\
    Abell 160 & 0.044 & 14.18 & 229 \\
    Abell 1644* & 0.047 & 14.80 & 403 \\
    Abell 780* & 0.054 & 15.19 & 243 \\
    CAN 40 & 0.035 & 14.09 & 223 \\
    Abell 3380 & 0.055 & 14.34 & 110 \\
    Abell 194* & 0.018 & 13.86 & 528 \\
    Abell 2589 & 0.042 & 14.55 & 334 \\
    Abell 2806 & 0.028 & 13.91 & 275 \\
    \textbf{Abell 2572a} & 0.042 & 14.53 & 16 \\
    Abell 1736 & 0.046 & 14.69 & 496 \\
    Abell 978 & 0.054 & 14.43 & 162 \\
    Abell 3736 & 0.049 & 13.65 & 162 \\
    Abell 3570 & 0.038 & 14.24 & 214 \\
    Abell 2665 & 0.056 & 14.56 & 171 \\
    Abell 1142 & 0.035 & 14.11 & 236 \\
    Abell 2657 & 0.040 & 14.58 & 313 \\
    HCG 62 & 0.015 & 13.92 & 739 \\
    \textbf{Abell 671} & 0.050 & 14.36 & 11 \\
    Abell S721 & 0.050 & 14.48 & 194 \\
    Abell S617 & 0.034 & 14.04 & 219 \\
    NGC1550 & 0.013 & 14.21 & 521 \\
    Abell S547 & 0.051 & 14.26 & 98 \\
    AM 0711 & 0.032 & 13.96 & 98 \\
    Abell 3565* & 0.012 & 14.33 & 733 \\
    Abell 536 & 0.040 & 14.96 & 132 \\
    Abell 3571* & 0.039 & 15.00 & 570 \\
    Abell 147* & 0.044 & 14.12 & 154 \\
    NGC1650 & 0.036 & 14.22 & 281 \\
    Abell 3990 & 0.029 & 14.93 & 139 \\
    Abell S141 & 0.019 & 13.83 & 223 \\
    Abell 3574E* & 0.016 & 14.50 & 642 \\
    Abell S851 & 0.010 & 13.91 & 624 \\
    Abell 3526 & 0.011 & 14.51 & 1566 \\
    Abell 4038 & 0.030 & 14.61 & 512 \\
    IC 1365 & 0.049 & 13.90 & 178 \\
    Abell 119* & 0.044 & 14.71 & 376 \\
    IC4992 & 0.019 & 13.95 & 333 \\
    Abell S805 & 0.015 & 13.87 & 569 \\
    \textbf{Abell S41} & 0.049 & 14.46 & 17 \\
    Abell 2593 & 0.043 & 14.44 & 315 \\
    Hydra (A1060)* & 0.013 & 14.40 & 1786 \\
    NGC1713 & 0.015 & 13.69 & 294 \\
    \end{tabular}
    \tablefoot{Clusters marked in bold indicate additional coverage added to clusters already included in the DR9 release. The columns and asterisks are as described in the Table\ref{tab:dr9_clusters} caption.}
    \label{tab:dr10_clusters}
\end{table}

\subsection{Catalogues and images}
We used images from the Dark Energy Camera Legacy Survey \citep[DECaLS:][]{Dey2019}, the same survey data utilised by the Galaxy Zoo: DESI/DECaLS project \citep{Walmsley2022}. The DECaLS survey uses the Dark Energy camera \citep[DECam:][]{Flaugher2015} on the 4m Blanco telescope at the Cerro Tololo Inter-American Observatory. DECam has a 0.32deg$^{2}$ field of view (FoV) with a plate scale of 0.262 arc seconds per pixel. The median point spread function has a full width at half maximum of 1."29, 1."18, and 1."11 in the g, r, and z bands, respectively.

The images have greater depth and higher resolution than the Sloan Digital Sky Survey \citep[SDSS, ][]{York2000} data used in the original iteration of Galaxy Zoo \citep{Lintott2008}, making them more suitable for identifying low-surface-brightness features in distant galaxies, as we aim to do with this project. For an initial release of 36 clusters for classification by the public, we utilised data from DECaLS DR9. We later added 43 clusters and additional coverage for three of the original clusters once DECaLS DR10 was available. We selected target galaxies from the DECaLS Tractor catalogue, using the Astro Data Lab ADQL query interface to impose initial selection criteria based on the Tractor morphologies, as described in Sect.\,\ref{sec:sampleselection}.

\subsection{Target clusters}

We selected 79 clusters across a redshift range of $0.0031 < z < 0.0562$, and X-ray mass range $13.3 < \logMhalo < 15.2$. to probe galaxy morphologies in a range of environments. The clusters were drawn from the X-ray selected catalogues of MCXC \citep{Piffaretti2011} and CODEX \citep{Finoguenov2020}.

We specifically included several clusters which overlap with the X-ray selected WINGS \citep{Fasano2006} and OmegaWINGS \citep{Gullieuszik2015} catalogues, since these surveys were inspected previously by \citet{Poggianti2016} to construct the GASP sample \citep{Poggianti2017,Poggianti2025}.
The GASP sample includes a MUSE large programme of confirmed ram-pressure stripped galaxies, with rich IFU data confirming the presence of offset ionised gas.
These clusters therefore provide an additional labelled dataset for validation and, in particular, an overlap with the GASP sample of ram-pressure stripped galaxies confirmed with integral-field spectroscopy.

Additionally, we selected several clusters that overlap with CHANCES \citep[the Chilean Cluster Galaxy Evolution Survey; ][]{Sifon2025}, which will provide additional spectroscopic coverage from the 4MOST Spectroscopic Survey Facility on the VISTA observatory.

Our full sample is representative of clusters across the halo mass range but undersamples lower redshift clusters ($\mathrm{z}\leq0.025$, see Fig. ~\ref{fig:z_dist}). We used the full sample for the subsequent analysis but show in Sect.\,\ref{sec:results} that using a representative subsample has no effect on our conclusions.

The final sample of DECaLS DR9 and DR10 clusters are shown in Tables\,\ref{tab:dr9_clusters} and \ref{tab:dr10_clusters} respectively. Clusters overlapping with the CHANCES sample are marked with asterisks.

\subsection{Sample selection}\label{sec:sampleselection}

A crucial factor for citizen science projects is user retention, i.e. how much time a given user is willing to contribute to completing classifications. A sense of meaningful contribution to scientific research has been shown to be the primary motivating factor for users to give more of their time and complete more classifications \citep{Raddick2013}. When selecting galaxies for the project, we therefore aimed to maximise the number of potential RPS candidates in the sample to ensure that a sufficient number of `interesting' candidates would appear during the classification process, thereby maintaining a user's interest.

We selected galaxies within a projected distance of $4\times R_{500}$ (corresponding to $2.87\times R_{200}$ on average) of their host cluster X-ray centres, since the majority of RPS interactions are expected to occur at low cluster-centric radii \citep{Jaffe2018}.
The choice of $R_{500}$ in the definition was motivated by the MCXC catalogue which defined only the $R_{500}$ radius of the clusters. At a later date, after the initial sample was live on the Zooniverse project, additional cluster parameters were made available from the CODEX catalogue \citep{Finoguenov2020} including the $R_{200}$ radius. We therefore maintained the $R_{500}$ constraint for consistency in the catalogue selection but used $R_{200}$ when defining cluster membership in the subsamples.
Of the full sample of 42684 objects, 9236 $(\sim20\%)$ have spectroscopic information, and we define a subsample of 2618 spectroscopically confirmed cluster members using the spectroscopic sample, selecting galaxies with $R<2\times R_{200}$ and $|\Delta V_{\rm LOS}|<4\sigma$.
We constrained the sample to galaxies which were classified by the DECaLS source extraction as \texttt{`EXP'} (exponential disk), \texttt{`REX'} (round exponential disk), and \texttt{`SER'} (modelled by a S\'ersic profile). For galaxies classified as \texttt{`SER'}, we further constrained the sample to a S\'ersic index of $n<2.5$ (including galaxies up to a S\'ersic index of 5 in a small number of cluster samples, for future comparison). Ram-pressure interactions have occasionally been reported in early-type galaxies \citep{Sheen2017}; however, the most common and most striking examples of ram-pressure disrupted galaxies are generally gas-rich, late-type galaxies \citep{Poggianti2016}, making them a prime target for maximising the expected fraction of RPS galaxies in the sample.

We cleaned the sample by imposing constraints using the DECaLS tractor catalogue bit-mask flags to remove sources which contain pixels flagged as saturated, hot pixels, bad columns or contaminated by cosmic rays. With the goal of cleaning the sample of galaxies that are too small or faint to reliably classify, we imposed a magnitude limit of 19 in the \textit{r} and \textit{g} bands and a minimum half-light radius of 2 arc seconds.

We downloaded red-green-blue (RGB) colour cutouts of the selected galaxies from the legacy survey image data using the default outputs, which show the low-surface brightness features of the jellyfish galaxies well. For each galaxy,  we auto-scaled the cutouts using the half-light radius to ensure that each galaxy filled the majority of the image. In addition, we downloaded a cutout with a $2.5\times$ wider FoV to show more of the galaxy's surroundings to better show extended tails and reveal any neighbouring or interacting companions that aren't visible in the narrow FoV cutout. 

\begin{figure}
    \centering
    \includegraphics[width=0.8\hsize]{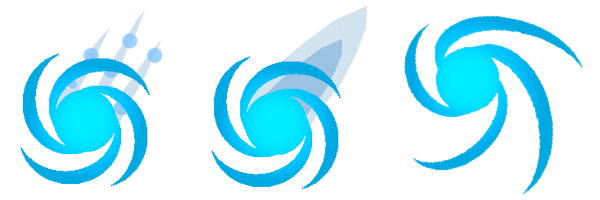}
    \includegraphics[width=0.7\hsize]{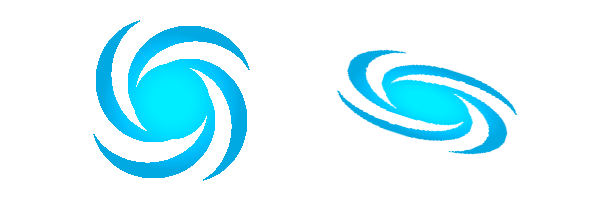}
    \includegraphics[width=0.8\hsize]{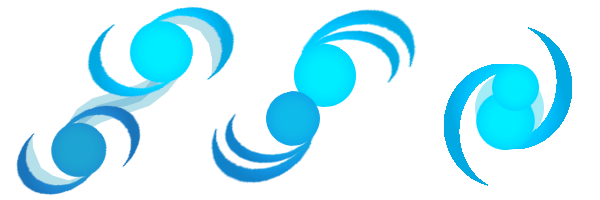}
    \caption{Examples of the icons used to denote various morphological features of the galaxies and provide a visual reference for the workflow options. \textit{Top row}: Disturbed, not merging. \textit{Middle row}: Undisturbed. \textit{Bottom row}: Disturbed, merging.}
    \label{fig:gzicons}
\end{figure}

\section{Classification process}\label{sec:classifications}

\subsection{Workflow}

The workflow of tasks, which are comprised of questions about physical characteristics of the subject, is shown in Fig~\ref{fig:workflow_diagram} of the appendix. The workflow was designed with the goal of getting as much information about the visual characteristics relevant to stripping whilst filtering out mergers and undisturbed galaxies, in as few steps as possible. The motivation for keeping the workflow as short as possible is to ensure that a volunteer is able to complete as many classifications as possible within a given time frame. Moreover, mergers and undisturbed galaxies should be identified as soon as possible so that the classifier can move onto the next galaxy, and minimal time is spent classifying galaxies we are not specifically searching for.

The workflow was designed to be as simple and descriptive as possible, so that users with little knowledge of the underlying science and little experience viewing images of disturbed galaxies would be able to classify galaxies based solely with reference to their appearance. To that end, visual icons describing different morphologies were included for most of the answers, showing different examples of features for users to classify based on physical features and comparisons. Several examples of the icons used are shown in ~Fig.~\ref{fig:gzicons}, with the full set of icons shown in the workflow diagram (see Fig~\ref{fig:workflow_diagram}) in the appendix.

The initial steps of the workflow were thus chosen to classify galaxies as either disturbed or undisturbed, and as merging or not merging, with `undisturbed' referring to galaxies with no signs of morphological disturbance, regardless of origin. Classifications of `undisturbed' or `merging' both end the workflow. If a galaxy was flagged as disturbed, but not merging, the classification process continued, focusing on the morphology of the galaxy and the nature of the disturbance.
The subsequent step in the workflow assessed the visibility of the spiral arms and, if present, whether they exhibit unwinding properties. In the subsequent step, we flagged the presence of a tail or extraplanar material. If a galaxy was classified with a tail, the final step in the workflow involved describing the tail's origin as either from the centre, elsewhere in the galaxy, or from the whole disk. The user was then prompted to draw the path of the tail using a line drawing annotation tool.

\begin{figure}
    \centering
    \includegraphics[width=\hsize]{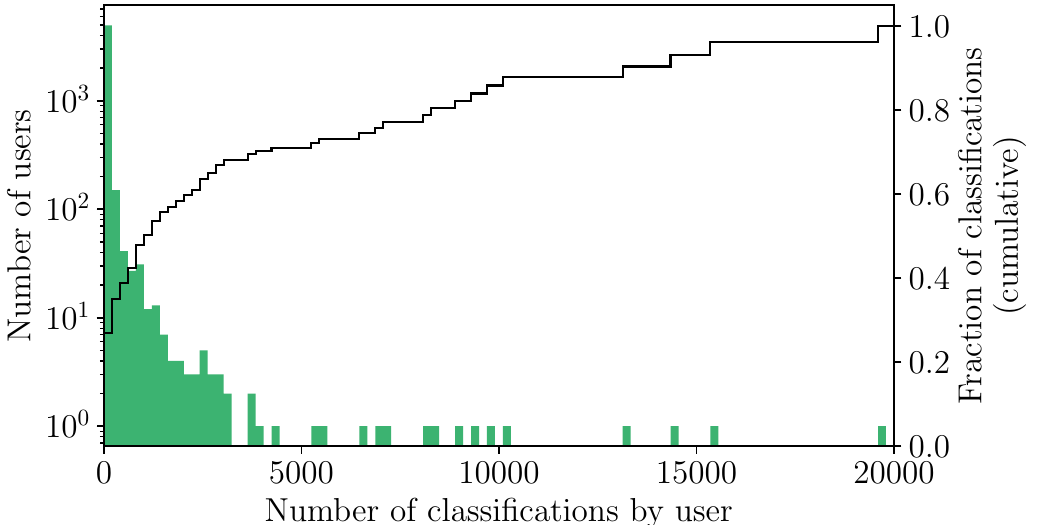}
    \caption{Green-shaded histogram showing the distribution of users according to the total number of classifications contributed.
    The black line indicates, for each bin, the cumulative total of classifications contributed by users in each bin.
    }
    \label{fig:user_classifications}
\end{figure}

\begin{figure}
    \centering
    \includegraphics[width=0.9\linewidth]{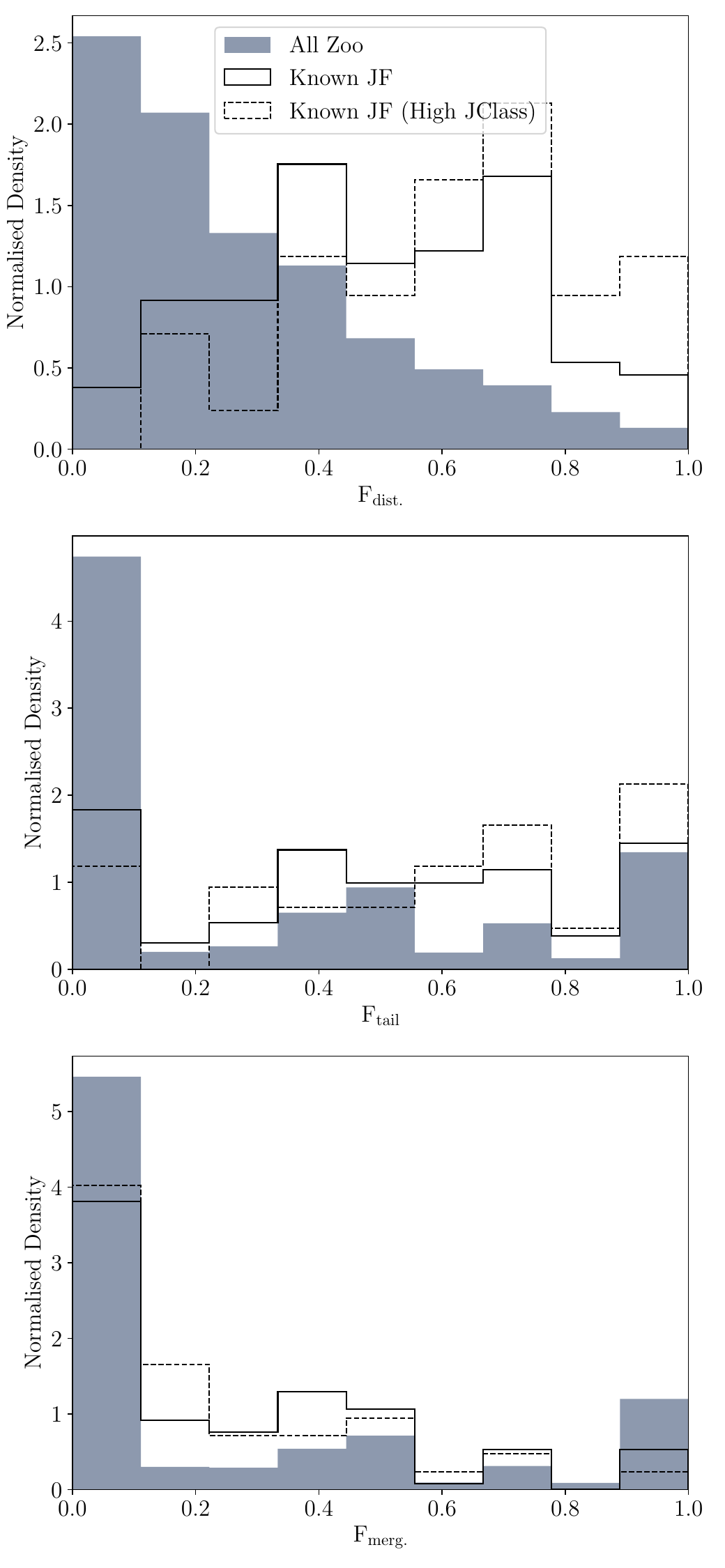}
    \caption{Histograms of the debiased vote fractions of galaxies matched with known ram-pressure stripped galaxies drawn from the literature. The grey-shaded histogram shows the full zoo sample. The black line shows all matches to known RPS galaxies (excluding those without optical tails) from the literature compiled in \citet{Crossett2025}. The dashed grey line shows the matches to the subset of known RPS galaxies classified as $\mathrm{JClass} \geq 3$, i.e. those with moderate to strong optical signs of stripping.}
    \label{fig:literature_galaxies}
\end{figure}

\subsection{Beta testing - A2626}
\label{sec:beta}

To validate and test the effectiveness of the classification process as well as verify the usability and clarity of the workflow for non-experts, we prepared a beta test, using a sample of galaxies using the same selection parameters as the main sample for the region within $4\times \mathrm{R}_{500}$ of A2626.
We also prepared tutorials and a field guide with information about galaxy clusters, the process of RPS, and merging interactions to give the volunteers context surrounding the galaxies and classification system.
For the beta test, we trialled different retirement limits (i.e. the number of classifications a subject must receive before it is no longer shown to classifiers) of 5, 10, and 20. A higher retirement limit provides more reliable classifications and dilutes the impact of erroneous classifications, whilst a lower limit allows a larger overall sample to be classified.

During the beta testing phase, we familiarised 400 non-expert volunteers classifying a sample of 290 galaxies with the Zooniverse platform. In total, we received 5829 classifications. To process the beta test results, we defined the sample using simplified classification criteria, assigning each object the label that received more than 50\% of the votes.

In particular, we defined `disturbed', `undisturbed', and `problem'  galaxies (representing cases hard to classify). The `disturbed' category was further subdivided into `merger' and `tail' galaxies, with the vote fractions  defined as

\begin{itemize}[itemindent=2cm]
    \item[``Disturbed'' -] \disturbedfrac{} = \disturbedfractioneqn{},
    \item[``Merger'' -]             \mergerfrac{} = \mergervotes{} / \disturbedvotes{},
    \item[``Tail'' -] \tailfrac{} = \tailvotes{} / \disturbedvotes{},
    \item[``Undisturbed'' -] \undisturbedfrac{} = \undisturbedfractioneqn{},
    \item[``Problem'' -] \problemfrac{} = \problemvotes{} / \totalvotes{},
\end{itemize}

where $\mathrm{V}_{\rm x}$ is the number of votes for a given feature x, and \totalvotes{} is the total number of classifications a galaxy has received.
These definitions were used for the vote fractions given in the catalogue and described in the subsequent sections.

Based on these criteria, we defined 230 undisturbed objects, 49 disturbed and not merging, seven disturbed and merging, and four galaxies flagged as a `problem'.

Based on visual inspection of the classified objects by experts, we determined that the public reliably identified disturbed morphologies but struggled to differentiate between merging and non-merging cases of disturbance, i.e. between gravitational and hydrodynamical (RPS) interactions, which are in fact hard to separate in many cases based on broadband imaging alone. To mitigate this difficulty, we updated the tutorial and field guide to include a wide variety of merging and stripped galaxies as a reference for the classifiers to compare.

To compare retirement limits, we calculated the classification of each subject after the first five votes, the first ten votes and the full set of 20 votes. We found that the majority of classifications did not change between each limit, with 10\% of classifications differing between five and 20 votes, and 4\% of classifications differing between ten and 20 votes. We therefore selected ten classifications as an optimal retirement limit for the main project. Assuming a fixed number of total classifications, this would double the sample of retired galaxies compared to 20 classifications, at a loss of only 4\% in accuracy.

\subsection{Results and vote fractions}\label{sec:fractions}

Approximately 18 months after launching the project on the Zooniverse platform, sufficient classifications were received to retire the full sample. In total, 5286 users (or unique user sessions for logged-out users) classified 49703 unique objects, with an average of 97.8 classifications per user. In total, 517,152 classifications were contributed, giving an average of 10.4 classifications per unique object. 
Figure~\ref{fig:classifications_hist} shows the distribution of classifications per image, while 
Fig.\,\ref{fig:user_classifications} shows the number of classifications completed by users as a red histogram. We found that 2197 users, approximately half of the total, classified under ten galaxies. On the other hand 667 users classified more than 100 galaxies. In the same figure, the black stepped histogram shows the total votes contributed by these users, revealing that whilst most users contributed fewer than 100 classifications, the small number of experienced users collectively contributed a significant fraction of the total. For the purpose of classification, we combined the votes for each subject, as explained in Sect.~\ref{sec:beta}, by defining vote fractions for each question in the workflow.

We examined galaxies in Abell 1644 to reevaluate the `problem' fraction, \problemfrac{}, defined as the proportion of citizens who considered the statement, 'The galaxy is not clear or too small / there is a problem with the image', true. 
Whilst we removed any sources flagged by the DECaLS bit masks, a small percentage of images still contained bright stars, missing bands, or other issues that could potentially hinder classification. The inclusion of the `problem' classification allowed volunteers to flag galaxies, which suffered sufficiently poor image quality to preclude classification.
Notably, however, we find no clear correlation between \problemfrac{} and image quality. 
Contrary to expectations, a high \problemfrac{} does not necessarily indicate an unclassifiable galaxy. In fact, low \problemfrac{} often correspond to poor image quality, while high \problemfrac{} may appear in galaxies that are clearly classifiable. This suggests that citizens use the `problem' label when uncertain about distinguishing between disturbed and undisturbed morphologies, rather than based solely on image quality. We speculate that the wording of the question in the workflow could have played a role in the misuse of the `problem' option, even though there were clear examples in the field guide. Since the sample was cleaned for images with saturation, cosmic rays, and bad pixels, galaxies -- which are genuinely unclassifiable due to issues with the images -- are therefore rare. In the rest of the analysis, we used a relaxed threshold, only flagging a galaxy as a `problem' only when $>80\%$ of users flag it as such.

\subsection{Classification criteria}\label{sec:criteria}

\begin{figure*}
    \centering
    \includegraphics[width=\linewidth]{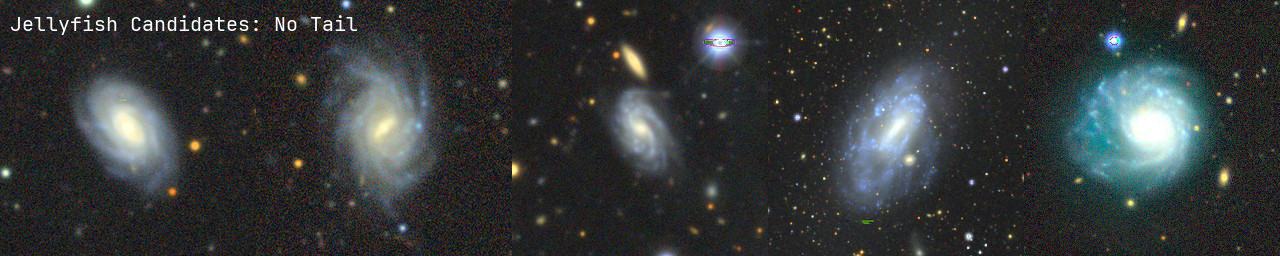}
    
    \includegraphics[width=\linewidth]{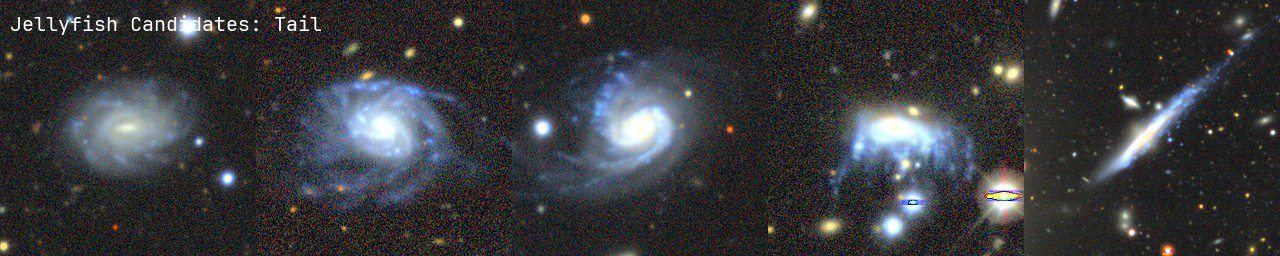}
    \caption{Selected examples of galaxies classified as SC based on the criteria described in the text. Top: Galaxies classified as having no tail-like feature. Bottom: Galaxies flagged as SC+T.}
    \label{fig:mosaics}
\end{figure*}

As a ground-truth for calibrating the selection of RPS candidates, a group of experts (co-authors of this paper) classified all 403 galaxies in the Abell 1644 cluster sample as mergers, jellyfish galaxies, and undisturbed galaxies. We scored the performance of the public classifications against the expert classifications using purity ($\mathcal{P}$) and completeness ($\mathcal{C}$). Purity is defined as the fraction of citizen-selected candidates that agree with the expert sample (Eq.~\ref{eq:purity}), and completeness is the fraction of expert-selected candidates recovered by the citizen sample (Eq.~\ref{eq:completeness}):

\begin{equation}
    \mathcal{P} = \frac{N^\circ_{\text{JF matched}}}{N^\circ_{\text{JF citizens}}},
    \label{eq:purity}
\end{equation}

\begin{equation}
    \mathcal{C} = \frac{N^\circ_{\text{JF matched}}}{N^\circ_{\text{JF experts}}},
    \label{eq:completeness}
\end{equation}

where $N^\circ_{\text{JF citizens}}$, $N^\circ_{\text{JF experts}}$, and $N^\circ_{\text{JF matched}}$ are the number of galaxies classified as jellyfish by the citizen science tester, expert classifiers, and both groups. A completeness of 100\% means all expert-selected candidates are included, while 100\% purity indicates all non-candidates are excluded.

We then explored a wide range of thresholds on the disturbed, non-merger, and tail vote fractions using a Monte Carlo approach. The optimal criteria for defining RPS galaxy candidates are

\begin{equation}
F_{\mathrm{dist.}} \geq 0.39\,\&\left(
F_{\mathrm{merg.}}\leq 0.23\,\|\,
F_{\mathrm{tail}} \geq 0.37 \right),
\end{equation}

which yield a calibrated selection with purity $P = 0.63$  and completeness $C = 0.57$. We defined galaxies meeting these criteria as SCs. The full performance landscape in the 3D parameter space is shown in Fig.,\ \ref{fig:Monte_carlo_plot} of the appendix. Whilst the purity level of $P = 0.63$ indicates that this sample is subject to contamination, we tested a stricter threshold of \disturbedfrac{}>0.5 and find no significant deviation from the results presented in the following section. Applying these thresholds to the Abell 1644 sample, we selected galaxies with morphologies consistent with jellyfish features, confirming that the chosen criteria provide a robust compromise between minimising contamination and recovering a large sample of ram-pressure stripped systems.

We defined undisturbed galaxies as those with \disturbedfrac{} < 0.1, referred hereafter as the reference sample (RS). We also included a sample of merging galaxies, defined as

\begin{equation}
F_{\mathrm{dist.}} \geq 0.39\,\&\,
F_{\mathrm{merg.}}\geq 0.23\,
\end{equation}

to which we refer hereafter as merging candidates (MCs).
Finally, within the definition of jellyfish candidates, we also defined a subsample exhibiting tail features as

\begin{equation}
F_{\mathrm{dist.}} \geq 0.39\,\&\,
F_{\mathrm{merg.}}\leq 0.23\,\&\,
F_{\mathrm{tail}} \geq 0.37.
\end{equation}

For the rest of the paper, we refer to these galaxies as SCs with tails (SC+T) and use this sample to illustrate the trends in stronger cases of stripping. This overlaps with the SC label but excludes cases of jellyfish-like disturbance where a significant tail is not present.

\begin{figure}
    \centering
    \includegraphics[width=\linewidth]{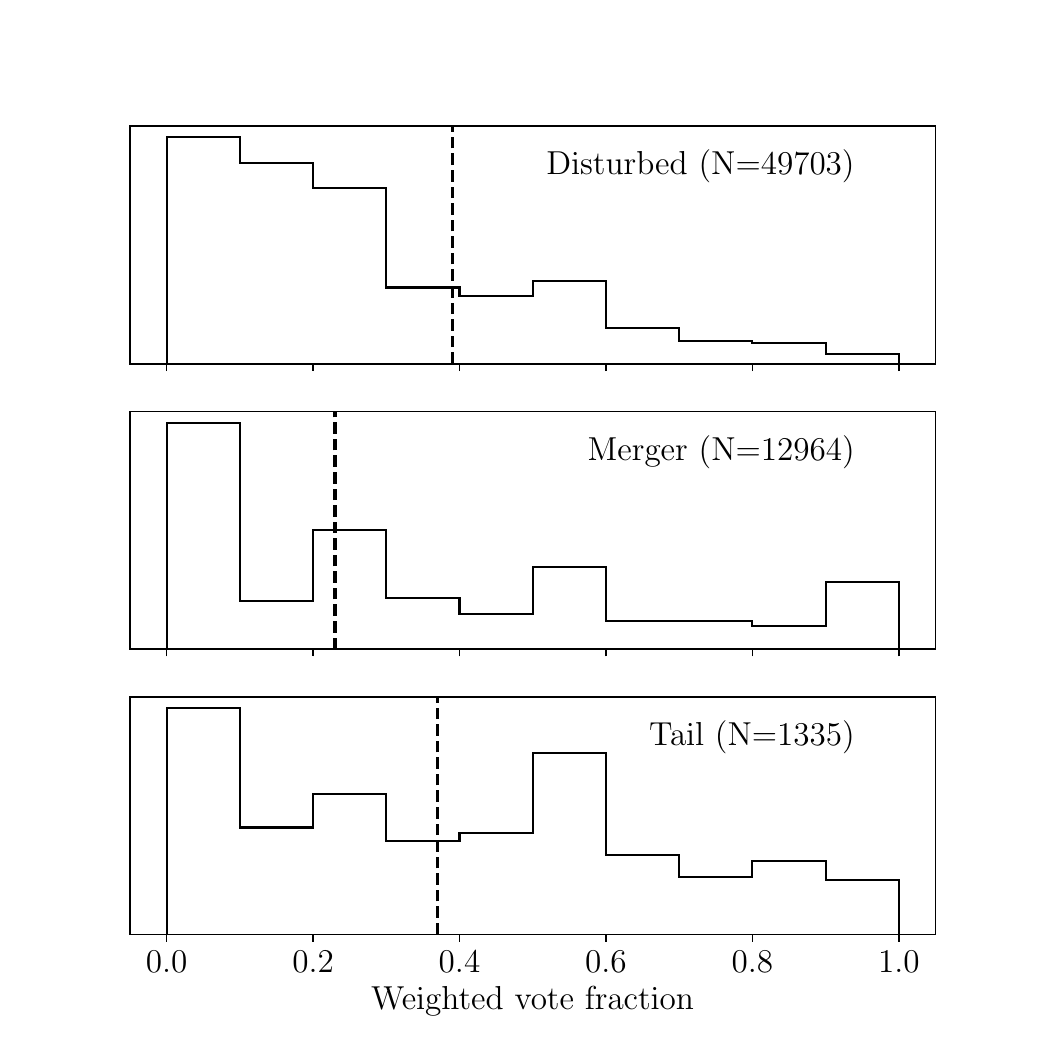}
    \caption{Weighted vote fractions for `disturbed', `merging', and `tail' labels, as defined in Sect.\,\ref{sec:combining}. The vertical dashed lines indicate our chosen criteria used to select galaxies within a specific sample. Middle: Galaxies defined as disturbed. Bottom: Significantly disturbed galaxies with \disturbedfrac > 0.8. The disturbed vote fraction shows a slight bimodality, which suggests that the classification step successfully distinguishes distinct populations within the sample. The merger vote fraction shows a peak at the lower end, suggesting a clear population of disturbed -- but not merging -- galaxies. Another smaller peak is present at the upper end, suggesting that the volunteers are in unilateral agreement in the clearest cases of merging.}
    \label{fig:vote_fractions}
\end{figure}

Figure~\ref{fig:vote_fractions} shows the thresholds for each classification criterion overlaid on the the corresponding distributions of vote fractions for the disturbed, merging, and tail labels across the sample.

\subsection{Combining and debiasing responses}\label{sec:combining}

\begin{figure}
    \centering
    \includegraphics[width=\hsize]{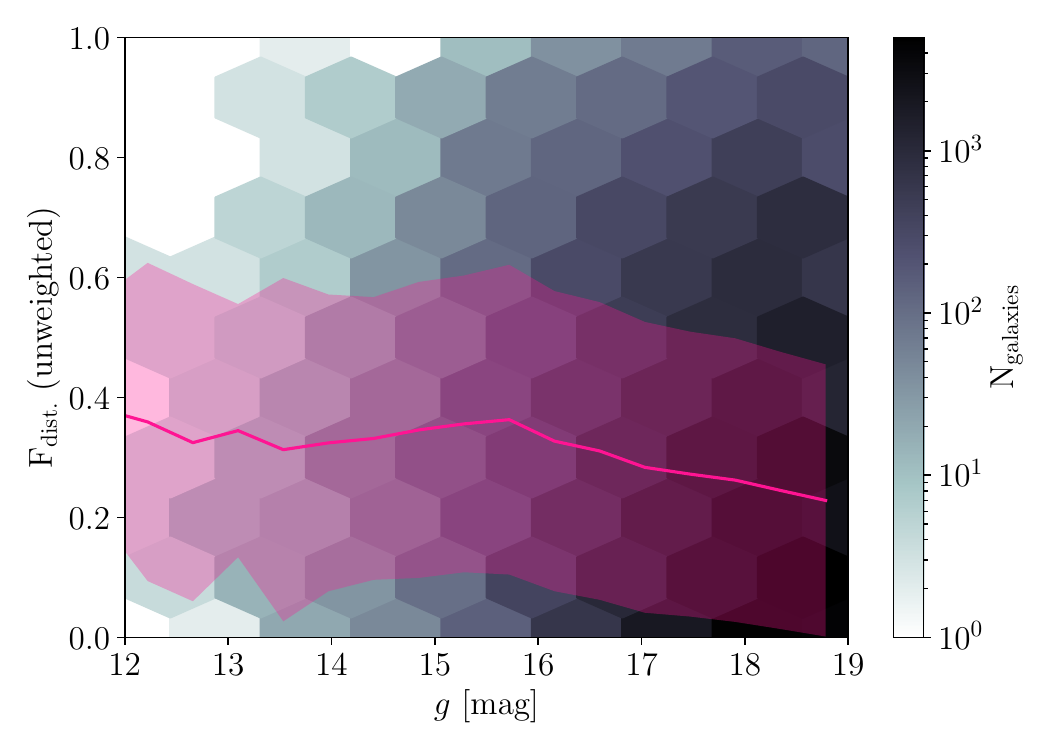}
    \caption{Distribution of \disturbedfrac{} vs g-band magnitude, for the full sample. The pink line shows the mean \disturbedfrac{}, binned by magnitude, whilst the shaded region indicates the standard deviation of \disturbedfrac{} within each bin. The grey-shaded hexagon shows the number of galaxies binned by magnitude and \disturbedfrac{}. A slight trend is visible with a smaller disturbed fraction at fainter magnitudes; however, the trend is within the overall scatter of the distribution of classifications in each magnitude bin.}
    \label{fig:classification_mag}
\end{figure}

An important step in producing a robust classification system is debiasing user votes. Debiasing the classifications homogenises the results across the pool of volunteers by prioritising patterns of votes consistent with the majority and minimises contamination by spurious voting patterns, well-intentioned or otherwise. There are multiple techniques for debiasing user scores, such as a weighting system that downweights the votes of users who classified $<N$ objects (i.e. less experienced users) or systems based on each user's concordance, i.e. the likelihood of a given user to agree with the majority.
Their voting behaviour was compared blindly, i.e. without comparing individual user votes to labelled data or known RPS galaxies.

We tested a combination of these systems to determine the optimal method for cleaning the vote distributions and minimising the impact of spurious or random votes. First, to `score' the classifications as predictors of the majority consensus, we used the fraction of user classifications in which the parameter matched the final classification. This was done separately for the following classifications: disturbed versus undisturbed, tail versus no tail, merging versus non-merging, and `problem' cases. Scores were allocated only when the final classification received more than two-thirds of the vote fraction (or 80\% for `problem' classifications), to avoid downweighting users in cases where the final classification was ambiguous. We explored this concordance metric against the number of classifications a user had completed, noting a minimum of ten classifications as the point at which a user typically had sufficient experience classifying galaxies and became a good predictor of the final outcome in at least 70\% of classifications.

In addition to a minimum threshold on the number of classifications, we investigated the concordance metric itself as a criterion for weighting user's votes. We trialled downweighting users with a high fraction of `disagreements', i.e. those who voted against other classifiers in more than $\mathrm{N}_{\rm thresh}$ classifications. $\mathrm{N}_{\rm thresh}$ is a tuneable parameter describing the maximum fraction of disagreements a user can have before being downweighted. This method of debiasing should, in principle, reduce the effect of spurious classifiers, such as users who misinterpret neighbouring galaxies as stripped material.

In previous galaxy zoo studies \citep[e.g.][]{Walmsley2022} debiasing was carried out using a brightness-limited sample, since faint objects are challenging to classify and debiasing them is unreliable. We investigated possible trends in the classifications: \disturbedfractioneqn{} against the g-band magnitude for the galaxies within our sample. When users struggled to identify disturbance in fainter galaxies, we expected a decreasing trend with magnitude. The results are shown in Fig.\,\ref{fig:classification_mag}. The figure indicates a slight trend, as expected, but this trend is well within the classification scatter in each magnitude bin. Since our sample is limited to a g-band magnitude of 19, it is unlikely that the objects are faint enough to cause issues with classifications; therefore, we did not apply an additional magnitude cut for the debiasing process. To test the effectiveness of the two debiasing methods, we compared the debiased and non-debiased user classifications on the pre-classified sample of Abell 1644 to measure the effect of debiasing on the completeness and purity of the user classified sample.

We find that debiasing the votes by downweighting users who classified fewer than ten objects by 50\% increased both the purity and completeness by 0.03 in cases where the vote thresholds were not optimised, i.e. the purity and completeness were lower; however, when the vote thresholds were tuned to maximise the completeness and purity, the difference is negligible.

On the other hand, debiasing the votes by downweighting users with a higher disagreement fraction yields slightly lower purity and completeness.
The observed reduction in purity and completeness likely result from the difficulty associated with identifying ram-pressure stripped galaxies. Users more proficient at identifying true ram-pressure-stripped galaxies (particularly in less obvious cases of RPS) are within the minority of the user-base and are therefore more likely to be downweighted according to their disagreement fraction.

Based on these findings, we opted to use debiasing only to downweight users who had carried out fewer than ten classifications, rather than basing it on the disagreement fraction. Although the gain from debiasing is negligible with the tuned vote thresholds, the debiased scores may be more useful in generalised cases, where different criteria may be applied to produce purer or more complete samples, as required.

We combined the debiased results using the same fractions defined in Sect.\,\ref{sec:fractions}, using the weighted rather than unweighted votes.

\subsection{Comparison with jellyfish galaxies from the literature}

To assess the effectiveness of the classifiers in recognising genuine cases of RPS, we spatially cross-matched the sample, based on optical position, with examples of ram-pressure stripped galaxies in the literature, i.e.\ those classified by experts in previous studies compiled from the catalogue in \citet{Crossett2025}. The images used in the literature classifications originate from different instruments, and the visibility of individual stripping signatures may vary; nevertheless, the general correlations against the literature classifications is an indicator of the effectiveness of the classification process.
Figure~\ref{fig:literature_galaxies} shows histograms of the debiased vote fractions of the `disturbed', `tail', and `merger' labels for the full sample of known RPS galaxies from the literature (excluding those exhibiting only radio or X-Ray tails)
and known RPS galaxies categorised by  \citet{Crossett2025} with a $\mathrm{JClass} \geq 3$ (i.e. with moderate to strong visual signs of stripping \citep{Poggianti2016}). The top panel shows a substantial difference in the distribution of the vote fraction between the full sample and the high-JClass matches, with the median \disturbedfrac{} approximately 1.5 times higher for the known jellyfish galaxies than for the full sample. This clearly indicates that the users are able to distinguish cases of disturbance due to RPS.
Moreover the high-JClass matches are skewed towards even higher vote fractions for disturbance, which confirms that the vote fraction correlates with the level of visual disturbance.
The middle panel shows that the jellyfish galaxies from the literature receive higher vote fractions for tail-like features, with high-JClass jellyfish galaxies skewed towards higher vote fractions. However, the difference is less pronounced than in the case of `disturbance' votes. The non-negligible population of literature jellyfish candidates with $F_{tail} \sim 0$ likely corresponds to galaxies with tails visible in deeper data, which are not as visible in the Legacy Survey images.
The lower panel of Fig.~\ref{fig:literature_galaxies} shows that, whilst the literature jellyfish galaxies are more frequently misidentified as merging galaxies than the full zoo sample, the overall vote fractions for merging are generally low. This suggests that users are not confusing RPS cases with merging cases. This is consistent with expert classifications. \citet{Poggianti2025} show that contamination by merging galaxies is very low ($\sim10\%$) in the visually identified sample of RPS candidates identified by \citet{Poggianti2016} in clusters, whilst \citet{Vulcani2021} show that contamination by non-RPS mechanisms increases to ($\sim70\%$) in the field.
Collectively, these plots reveal that true jellyfish galaxies confirmed in the literature receive more votes for disturbed morphologies, indicating that the public are successfully able to distinguish genuine visual signs of stripping.

\section{Results}\label{sec:results}

In Sect.~\ref{sec:catalogue} we present the catalogue of RPS candidates generated by this project. In the sections that follow, we discuss using this catalogue to test the effectiveness of the public classification method in selecting galaxies undergoing physical interactions. We examine the expected properties of the SC, SC+T, and MC samples, including their locations within clusters, morphologies, colours, and SFRs. The classifiers were provided no information on the physical parameters. Correlations with phase-space position and star-formation rate, which are stronger in the classified SC and SC+T samples than in the undisturbed sample, can therefore be interpreted as confirmation that the visually classified samples experience hydrodynamical interactions. To incorporate stellar masses and star-formation rates into our analysis, we cross-matched our classified galaxies with SDSS using a 3-arc-second threshold, retrieving derived properties for 9816 matched sources from the MPA-JHU emission-line analysis of SDSS DR7.

\subsection{Catalogue of ram-pressure stripped candidates}\label{sec:catalogue}

Based on the criteria derived in Sect.\,\ref{sec:criteria}, we obtained a catalogue of 6739 SC galaxies, of which 3910 exhibit tail features (SC+T), 5430 MC galaxies, and 29729 RS galaxies. Figure~\ref{fig:mosaics} shows examples of SC and SC+T galaxies.

The initial catalogue of classified objects released along with this paper consists of 37,599 objects from the clusters sample. An extract from the catalogue is shown in Table~\ref{table:catalogue} of the appendix. The catalogue contains the right ascension and declination of each galaxy derived from the DECaLS tractor catalogue, along with the total number of classifications, weighted and unweighted vote fractions for each feature, and weighted and unweighted vote counts for each feature. The weighted votes and fractions were derived from the debiasing process outlined in Sect.\,\ref{sec:combining}. The fractions are defined in Sect.\,\ref{sec:fractions}. The additional group sample and Shapley supercluster sample will be released alongside forthcoming papers in this series.



\subsection{The abundance of RPS candidates across environments}

\begin{figure}
    \centering
    \includegraphics[width=\linewidth]{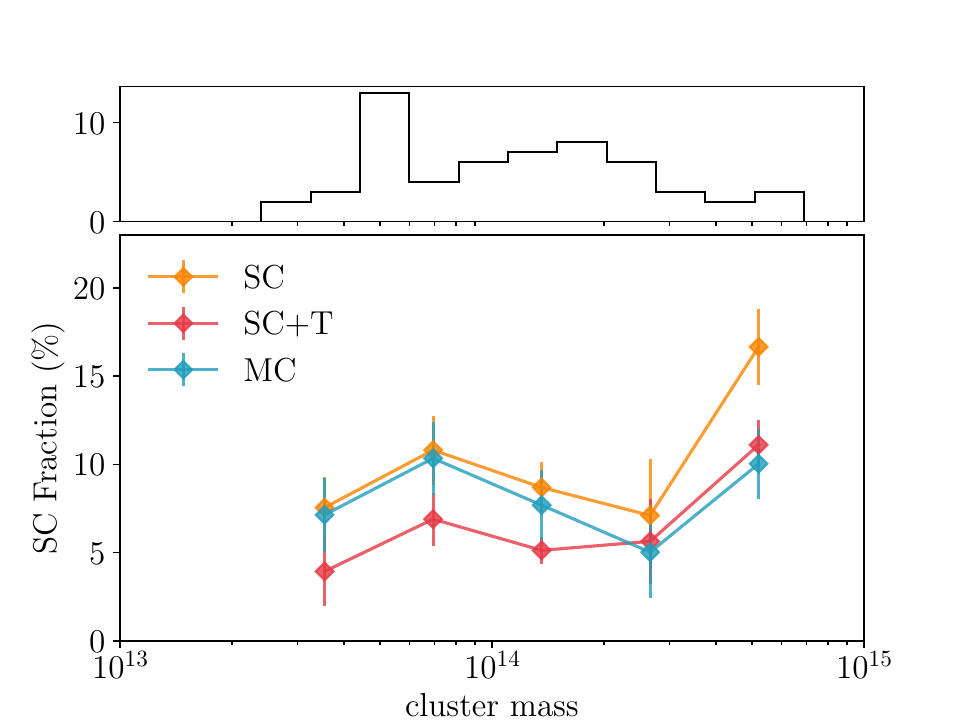}
    \caption{Fraction of galaxies classified as SC (orange), MC (blue), and SC+T (pink) by the volunteers, shown as a function of host cluster or group mass. The error bars indicate the standard error within each mass bin. The histogram shows the number of clusters across the halo mass range of the sample.}
    \label{fig:halomass_fraction}
\end{figure}

\begin{figure}
    \centering
    \includegraphics[width=\linewidth]{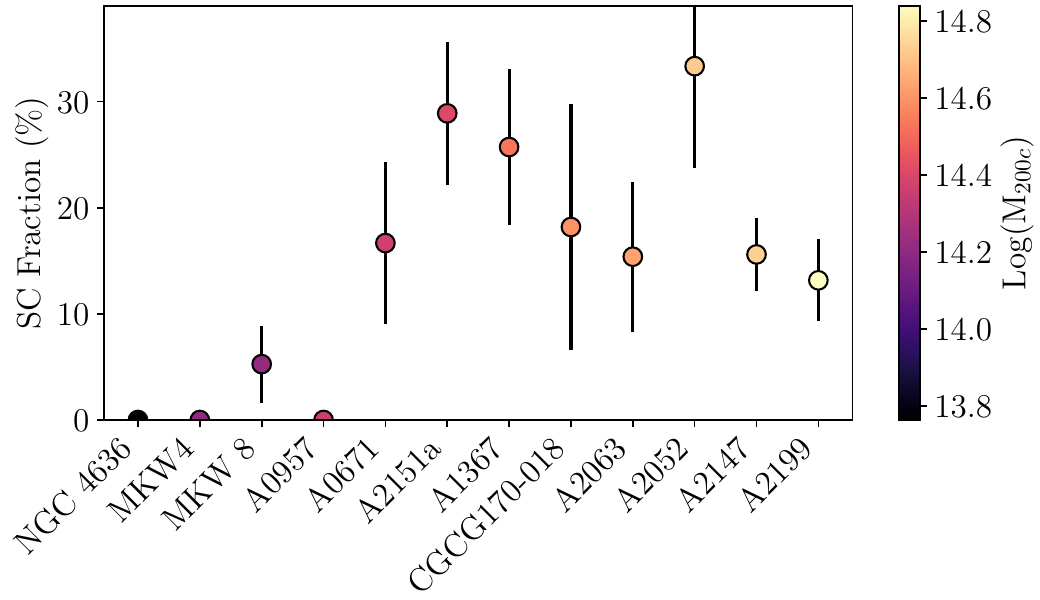}
    \caption{Fraction of jellyfish candidates over all star-forming galaxies in clusters with spectroscopy and SDSS overlap, showing only spectroscopically confirmed members and galaxies with $M_*>10^{9.7}\mathrm{M}_\odot$. The error bars show the $1\sigma$ binomial errors following a normal approximation.}
    \label{fig:fraction_clusters}
\end{figure}

One of the key open questions that studies such as this one hope to answer is how important RPS is for galaxy transformations in clusters \citep{Sampaio2022,Sampaio2024}. This is a challenging question to answer due to observational limitations, and optical studies such as ours only observe strong cases of stripping where star formation occurs in the stripped debris. However, it has been shown that these optically selected samples, while incomplete, are pure \citep[at an $>$80\% level for galaxies in clusters][]{Poggianti2025}. Therefore, quantifying the fraction of optically selected RPS candidates in clusters sets a robust lower limit on the influence of RPS. 

Figure~\ref{fig:halomass_fraction} shows the fraction of jellyfish candidates, mergers, and jellyfish candidates with tails as a function of cluster halo mass for spectroscopically confirmed members. The fractions are defined as the number of galaxies in each sample divided by the total number of classified subjects within the cluster; therefore, both the numerator and denominator are both subject to the same selection criteria defined in Sect.\,\ref{sec:sampleselection}. The top panel shows a histogram of the cluster halo masses across the sample. The plot shows that both the SC and SC+T samples exhibit an increasing fraction with cluster halo mass, whereas the merging galaxies exhibit a flatter, fluctuating trend with host cluster mass.

Figure~\ref{fig:halomass_fraction} shows that the fraction of galaxies classified as SC is generally independent of halo mass for clusters below $3\times10^{14}\mathrm{M}_\odot$. Above this value, the SC fraction increases sharply. With more higher-velocity infallers and a denser ICM, it is naturally expected that more massive clusters host environments more conducive to RPS. A similar trend with halo mass is observed in \citet{Roberts2021,Roberts2021b}.

The sample of MC galaxies fluctuates in their observed fraction as a function of environment, exhibiting the same overall shape as the SC sample with a less significant increase at higher halo masses. 
Due to dynamical friction, a lower merger rate in clusters is expected compared to groups, owing to the high velocity dispersion observed in the former. Yet, galaxies in clusters can still face high-speed encounters, known as harassment \citep{Moore1996} that may also affect their morphology. 
Since our classification images do not reveal the end state of the galaxies' interactions, merging galaxies and those undergoing harassment are challenging to visually distinguish, and it is likely that a combination of these two processes drives the observed fluctuation in fraction across cluster mass.

Figure~\ref{fig:fraction_clusters} shows the fraction of SC galaxies in a sample of clusters with both spectroscopic data and SDSS overlap, with clusters ordered by halo mass. Stripping candidate fractions were calculated relative to star-forming galaxies, and a stellar mass limit of $M_*>10^{9.7}\mathrm{M}_\odot$ (derived from the SDSS MPA-JHU catalogue), was imposed, following the methodology of \citep{Roberts2021}, to ensure stellar mass completeness. The error bars show the $1\sigma$ binomial errors on the fraction calculated following the normal approximation described in \citet{Cameron2011}. The plot shows a trend similar to that in Fig.\,\ref{fig:halomass_fraction}, with clusters below $\sim10^{14.3}\mathrm{M}_\odot$ exhibiting low fractions. Above this threshold, the plot reveals that the SC fraction is largely independent of environment, with a maximum of around $30\%$ of star-forming galaxies exhibiting visual signs of RPS in A2151a and A2052. A2052 exhibits an excess in the fraction of RPS candidates compared to clusters of similar mass. This is likely due to sloshing processes resulting from the cluster's highly disturbed nature \citep{Blanton2011}, which can boost stripping events \citep{Roediger2014,Stroe2015,Stroe2020,Bellhouse2022}. With the exception of A2052, clusters in the middle of the mass range exhibit higher jellyfish fractions than those at the upper end of the range.

As discussed in Sect.\,\ref{sec:sample}, our sample of clusters is representative across the halo mass range at the 5\% significance level (p-value 0.20); however, the full sample undersamples clusters at low redshifts ($\mathrm{z}\leq$0.025). If we restrict our clusters to redshifts above this limit, the resulting sample is representative across both halo mass and redshift at the 5\% significance level, and the observed trends with halo mass remain unchanged.

\subsection{Location within the clusters}

\begin{figure}
    \centering
    \includegraphics[width=\linewidth]{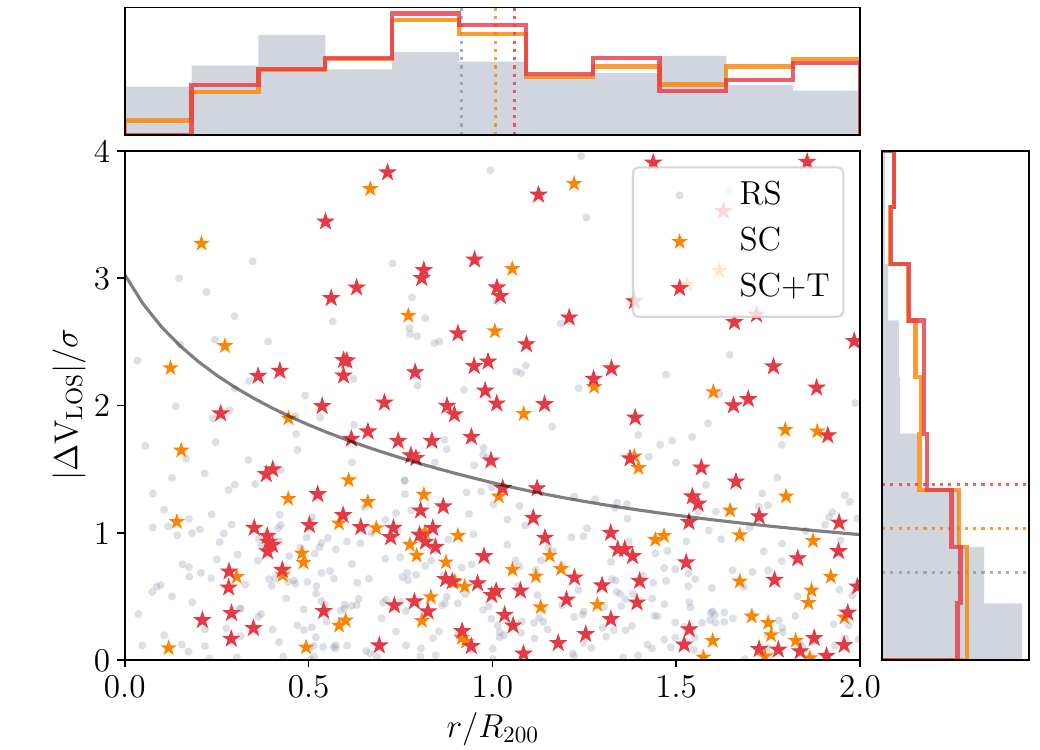}
    \caption{
    Projected phase-space diagram for galaxies spectroscopically confirmed as cluster members, separated into RS (grey dots), SC (orange stars), and SC+T (red stars). Middle panel: Escape velocity (grey line) of an NFW profile with a concentration parameter, $c=6$. Top and right panels: Radial and velocity distributions, respectively. In the histograms, we also show the median value for each subgroup.
    }
    \label{fig:phasespace}
\end{figure}

To explore whether the publicly classified SC and SC+T galaxies exhibit any trends or tendencies with location in their host clusters, we measured the phase-space locations of the subset of galaxies with spectroscopic redshifts, restricting the samples to show only galaxies within $2\times\mathrm{R}_{200}$ of their host cluster centres with $|\Delta V_{\rm LOS}| / \sigma_{\rm cluster} < 4$. Figure\,\ref{fig:phasespace} shows the phase-space plot, along with histograms of the cluster-centric radius and velocity, for the RS, SC, and SC+T samples. The histograms are normalised to show the density of each sample, to facilitate comparison. The individual galaxies are shown in the phase-space plot as grey, orange, and blue points for the undisturbed, jellyfish candidate, and merger samples, respectively.
The curves show the escape velocity calculated for an Navarro-Frenk-White \citep[NFW][]{NFW} profile of a cluster with a concentration parameter of $c=6$. The dashed line indicates the completeness limit, constrained by the $4\times\mathrm{R}_{500}$ criterion used to select the initial catalogue. The mean limiting radius is $2.87 \times\mathrm{R}_{200}$ across the full sample of clusters; however, for the most centrally concentrated cluster, the cutoff value of $4\times\mathrm{R}_{500}$ limits the selection to $\sim 2 \times\mathrm{R}_{200}$.

The velocity histogram shows that the distributions of the SC and SC+T samples shift slightly towards higher absolute line-of-sight (LOS) velocities compared to the undisturbed sample by $\sim$20\%.
Kolmogorov–Smirnov (K-S) tests of the distributions of the velocities of each sample reveal that they are distinct from the distribution of undisturbed galaxies at the 5\% level, with p-values of 0.0032 (SC) and 0.0007 (SC+T).

Interestingly, the SC galaxies are found at larger cluster-centric radii than undisturbed galaxies, but only by $\lesssim$10\% on average. A K-S test shows that the distributions of cluster-centric radii of both the SC sample and the SC+T sample are distinct from the RS galaxies at the 5\% level, with p-values of 0.007 and 0.021 respectively.

Since ram-pressure strength scales with galaxy velocity and environmental density, jellyfish galaxies are typically found at higher LOS velocities in phase space and at lower cluster-centric radii \citep{Jaffe2015, Rhee2017, Jaffe2018}. This is also an indication that they are on first infall at preferentially radial orbits \citep{Salinas2024,Biviano2024}. Our sample shows that the selected SC galaxies, and to a lesser extent the MC galaxies, are kinematically distinct from the RS galaxies, as expected for recent infallers.

\subsection{Galaxy colours and morphologies}

\begin{figure}
    \centering
    \includegraphics[width=\linewidth]{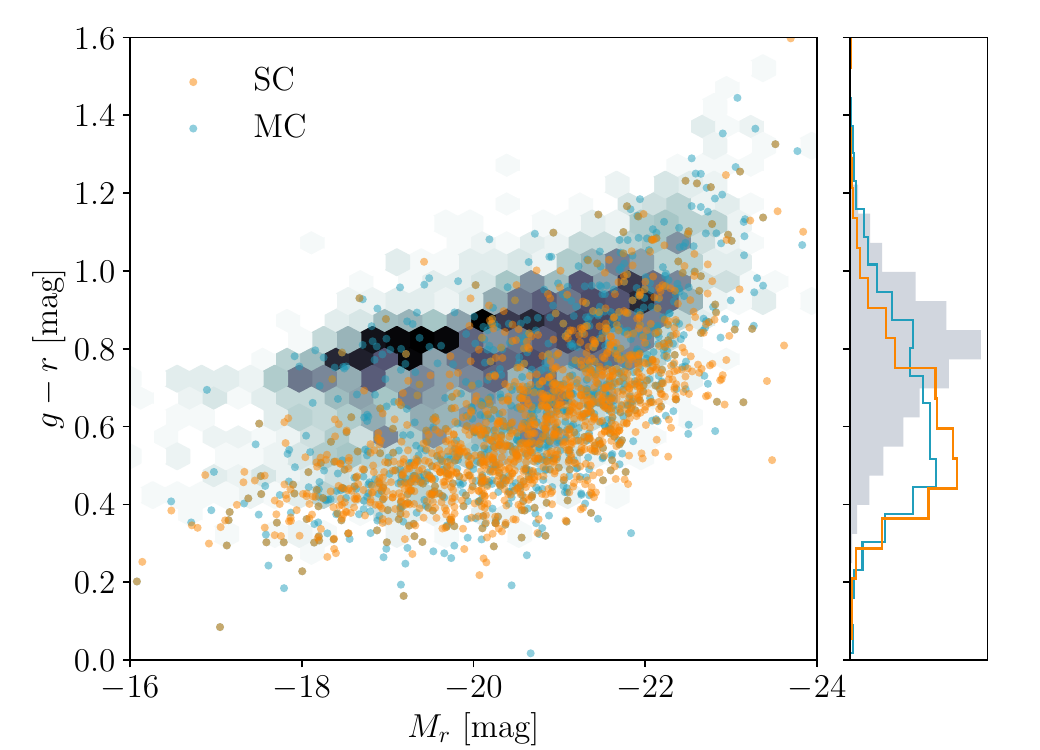}
    \caption{Colour-magnitude diagram of the full RS, SC, and MC samples, showing the full spectroscopic samples, i.e. not limited to galaxies with cluster membership. The undisturbed sample is shown as a greyscale density map, whilst the jellyfish and merger samples are shown as individual points. The histogram shows the distribution of g-r colours for the three samples.}
    \label{fig:colormag}
\end{figure}
\begin{figure}
    \centering
    \includegraphics[width=\linewidth]{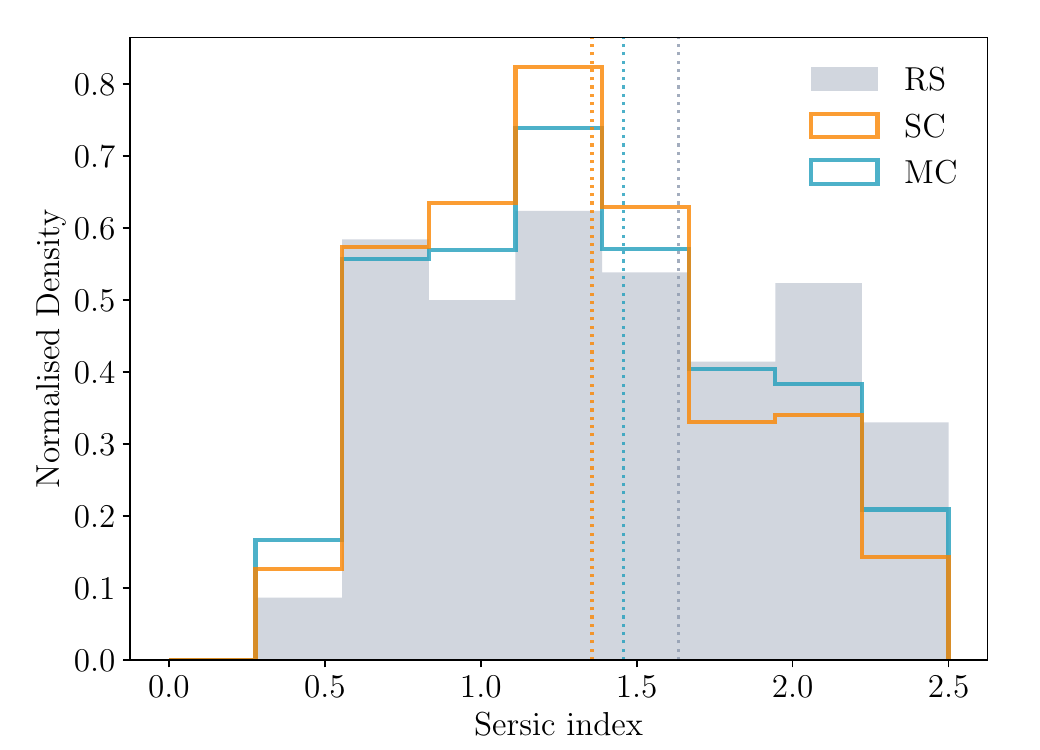}
    \caption{Histogram of S\'ersic index for the RS, MC, and SC samples, showing all galaxies classified as \texttt{`SER'} in the DECaLS tractor catalogue, including objects without spectroscopy.}
    \label{fig:sersic-hist}
\end{figure}

We also compared the colours and morphologies of the classified samples. Whilst the magnitude limits of the sample selection implicitly impose a colour cut, the selection criteria did not explicitly limit the sample to blue galaxies. Instead, the sample is constrained by the morphology criteria described in Sect.\,\ref{sec:sampleselection}. Figure~\ref{fig:colormag} shows the colour-magnitude diagram for the spectroscopic sample, not limited by cluster membership, with the histogram of colours shown in the panel to the right of the figure. The grey hexbin plot shows the RS galaxies, which show significantly redder g-r colours compared to both of the SC and MS samples. The MC sample, shown by the blue points, are biased towards the blue cloud but are present in both regions. The SC galaxies, shown as orange points, clearly favour the blue cloud. This is evidence that the publicly identified SCs tend to have young stellar populations, in comparison with the overall sample. The histogram shows that both the SC and MC samples are significantly bluer than the RS galaxies. Moreover, K-S tests reveal that both populations are distinct from the RS galaxies with p-values consistent with zero in both cases. This is consistent with the findings of \citet{Vulcani2022} who compared the B-V colours of both SCs with a reference sample of cluster galaxies.

In terms of morphologies, both the MC and SC samples have late-type morphologies, and the subsamples of galaxies classified as \texttt{`SER'} in the DECaLS tractor catalogue exhibit slightly lower S\'ersic indices on average compared to the RS galaxies, as shown in Fig.~\ref{fig:sersic-hist}. Whilst the average S\'ersic indices are consistent across the samples, K-S tests show that the distributions of the S\'ersic index in both the SC and MC samples differ from the RS galaxies, with p-values consistent with zero. Both the SC and MC samples exhibit slight biases towards lower S\'ersic indices, which is consistent with \citet{Vulcani2022} who found that visually identified SCs generally exhibit late-type morphologies compared with the reference sample. Nevertheless, our morphology constraints in the initial sample selection yield a lesser distinction between the populations, as our sample was pre-selected to include late type morphologies.

\subsection{Star formation rate}

\begin{figure*}
    \centering
    \includegraphics[width=\linewidth]{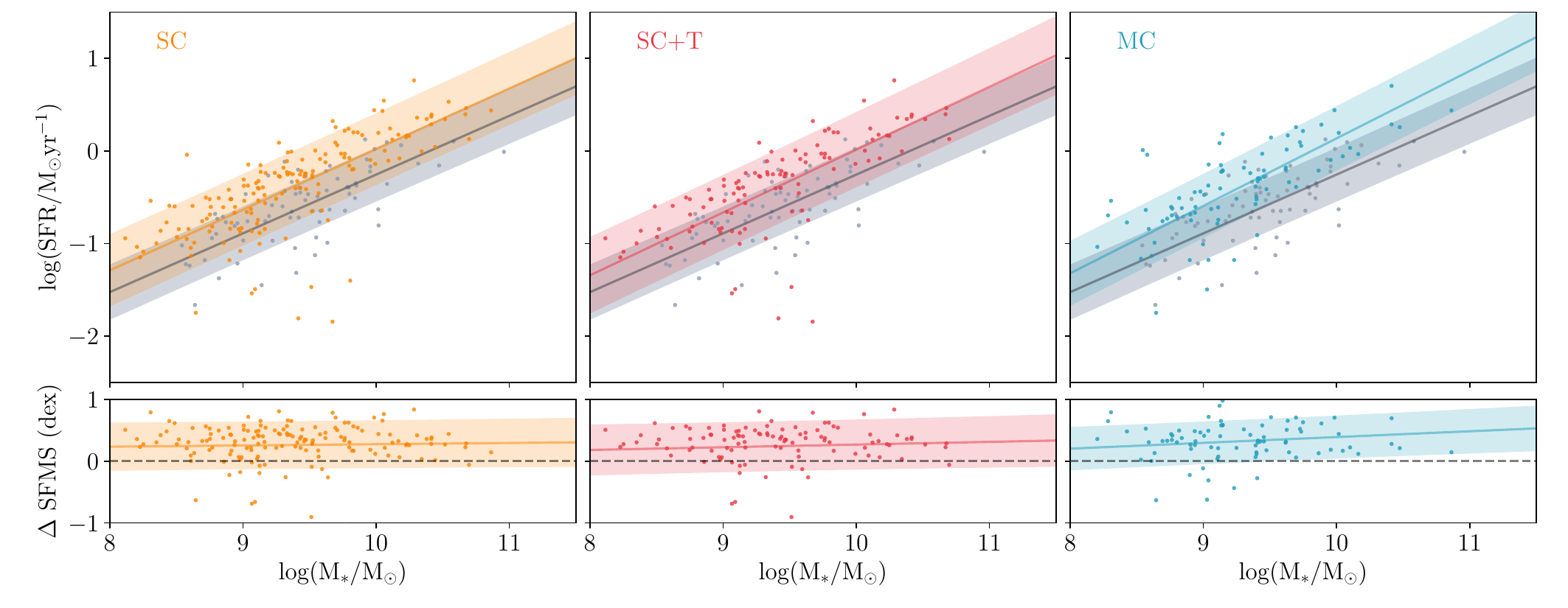}
    \caption{SFR mass relation for galaxies in the SC, SC+T, and MC samples, using the full classified sample of galaxies cross-matched with SDSS. The coloured points show individual galaxies in the sample, whilst the grey hex-plot shows the density of the RS galaxies. The coloured lines and shaded regions indicate a linear fit and 1$\sigma$ confidence interval respectively.}
    \label{fig:sfr}
\end{figure*}

To investigate whether the SC, SC+T, and MC samples exhibit enhanced SFR in comparison to other blue galaxies, we performed a more detailed analysis of the SFR across the samples using the SDSS cross-matched subsets. Jellyfish galaxies are typically characterised as star-forming or composite systems \citep{Poggianti2017b,Roberts2021}. The MPA-JHU catalogue assigns a \texttt{bptclass} parameter, which distinguishes the dominant ionisation sources according to the Baldwin, Phillips, and Terlevich diagram \citep[BPT;][]{Baldwin1981}. This parameter classifies each source as one of six categories: (i) star-forming, (ii) low-SN star-forming, (iii) composite, (iv) Seyfert, (v) LINER, and (vi) unclassified. To mitigate contamination from galaxies with only weak nebular emission, we required an $\rm H\alpha$ equivalent width $EW(H\alpha) > 3$\,\AA, consistent with the WHAN diagnostic threshold used to distinguish genuine LINERs from `retired' galaxies powered by old stellar populations \citep{CidFernandes2010}. This ensures that systems dominated by faint emission from evolved stars are excluded from our analysis. An estimate of the uncertainties in the emission lines can be seen in Table~2 of \citet{Sampaio2023}. We selected only objects identified as star-forming in the BPT diagram, for which SFRs were computed directly from the $\rm H\alpha$ emission-line flux.

Figure~\ref{fig:sfr} shows the SFR versus stellar mass relation for the SC, SC+T, and MC samples. In this figure, the full sample of classified, cross-matched galaxies is shown with no constraints on cluster membership. The figure shows the undisturbed sample as a grey hexbin plot, with the individual galaxies in the SC, SC+T, and MC samples coloured in orange, pink, and blue, respectively. The solid lines and shaded regions show the linear fits and $68\%$ confidence intervals of each sample.

The figure shows that all samples exhibit elevated SFR in comparison to the undisturbed RS galaxies, with the difference being more pronounced at lower stellar masses. Ram-pressure stripping is expected to produce an elevated SFR, including in visually identified samples \citep{Poggianti2016,Vulcani2020b,Roberts2020,Vulcani2022}. Our results confirm that the optically selected SC galaxies selected by the public exhibit a tentative excess in SFR and, therefore, are consistent with results for galaxies undergoing environmentally driven disruption as found in the literature.

\section{Conclusions}\label{sec:conclusions}

We presented the first results of \textit{Fishing for Jellyfish Galaxies}, 
the inaugural citizen-science project aimed at identifying galaxies undergoing 
RPS in optical data. Using imaging from the Dark Energy Camera Legacy Survey 
and a tailored workflow on the Zooniverse platform, more than 4400 volunteers 
classified nearly 50,000 galaxies across 82 clusters.  

By calibrating vote-fraction thresholds against expert classifications in one cluster (Abell~1644), we defined robust selection criteria with a completeness of 0.57 and Purity of 0.63 that yield a catalogue of 
6739 jellyfish candidates (SC), 5430 mergers (MC), and 29729 undisturbed galaxies (RS). A subset of 3910 candidates exhibit clear tail-like features (SC+T), confirming their morphological similarity to previously identified jellyfish systems.

An initial analysis shows that SC galaxies preferentially occupy larger 
cluster-centric radii and higher line-of-sight velocities than undisturbed galaxies, 
consistent with conditions where RPS is most effective. They also display enhanced 
SFRs, as reported by previous studies, providing independent and more statistically robust evidence that the selected galaxies experience hydrodynamical interactions. An analysis of the spectroscopically confirmed SC galaxies and their host clusters confirms a trend with environment, with more massive clusters hosting a larger fraction of SC and SC+T galaxies, as well as an increased fraction of MC galaxies.

These results demonstrate that citizen scientists can reliably identify galaxies 
affected by environmental processes, even when the signatures are subtle. The 
resulting catalogue of $\sim$49,703 visually classified galaxies represents the 
largest homogeneous sample of its kind to date and provides a valuable resource 
for future studies of galaxy transformation in clusters.

Looking ahead, this catalogue will serve as a training set to fine-tune automated 
methods for detecting RPS candidates in large surveys (Jaff\'e et al., in prep.), 
and to guide the design of future citizen-science projects aimed at identifying galaxies undergoing a variety of environmental processes (e.g. 'Looking for strange galaxies'\footnote{\url{https://www.zooniverse.org/projects/vitorms/looking-for-strange-galaxies}}). It will also enable 
statistical studies of the detailed properties of RPS candidates in clusters, an investigation of the `unwinding' galaxy population. In a forthcoming publication, we will further present the extension of the `Fishing for Jellyfish' project that uses deeper optical data from HSC 
and a more-detailed flowchart for classifications (Quiroz et al. in prep.).

\section*{Data availability}

The catalogue of classified galaxies produced by this project is only available in electronic form at the CDS via anonymous ftp to cdsarc.u-strasbg.fr (130.79.128.5) or via \href{http://cdsweb.u-strasbg.fr/cgi-bin/qcat?J/A+A/}{http://cdsweb.u-strasbg.fr/cgi-bin/qcat?J/A+A/}.

\begin{acknowledgements}

This publication uses data generated via the Zooniverse.org platform, development of which is funded by generous support, including a Global Impact Award from Google, and by a grant from the Alfred P. Sloan Foundation. We gratefully acknowledge the thousands of volunteers who participated in the Fishing for Jellyfish Galaxies Zooniverse project. Their classifications made this work possible. We thank the Zooniverse community for their enthusiasm, careful visual inspections, and valuable discussions on the Talk forums.

This project has received funding from the European Research Council (ERC) under the European Union's Horizon 2020 research and innovation programme (grant agreement No. 833824).

YLJ acknowledges support from the Agencia Nacional de Investigaci\'on y Desarrollo (ANID) through Basal project FB210003, FONDECYT Regular projects 1241426 and 123044, and  Millennium 
Science Initiative Program NCN2024\_112.

\end{acknowledgements}

\bibliographystyle{aa}
\bibliography{ref}

@ARTICLE{Bamford2009,
       author = {{Bamford}, Steven P. and {Nichol}, Robert C. and {Baldry}, Ivan K. and {Land}, Kate and {Lintott}, Chris J. and {Schawinski}, Kevin and {Slosar}, An{\v{z}}e and {Szalay}, Alexander S. and {Thomas}, Daniel and {Torki}, Mehri and {Andreescu}, Dan and {Edmondson}, Edward M. and {Miller}, Christopher J. and {Murray}, Phil and {Raddick}, M. Jordan and {Vandenberg}, Jan},
        title = "{Galaxy Zoo: the dependence of morphology and colour on environment*}",
      journal = {\mnras},
         year = 2009,
        month = mar,
       volume = {393},
       number = {4},
        pages = {1324-1352},
          doi = {10.1111/j.1365-2966.2008.14252.x},
archivePrefix = {arXiv},
       eprint = {0805.2612},
 primaryClass = {astro-ph},
       adsurl = {https://ui.adsabs.harvard.edu/abs/2009MNRAS.393.1324B}
}

@ARTICLE{Willett2013,
       author = {{Willett}, Kyle W. and {Lintott}, Chris J. and {Bamford}, Steven P. and {Masters}, Karen L. and {Simmons}, Brooke D. and {Casteels}, Kevin R.~V. and {Edmondson}, Edward M. and {Fortson}, Lucy F. and {Kaviraj}, Sugata and {Keel}, William C. and {Melvin}, Thomas and {Nichol}, Robert C. and {Raddick}, M. Jordan and {Schawinski}, Kevin and {Simpson}, Robert J. and {Skibba}, Ramin A. and {Smith}, Arfon M. and {Thomas}, Daniel},
        title = "{Galaxy Zoo 2: detailed morphological classifications for 304 122 galaxies from the Sloan Digital Sky Survey}",
      journal = {\mnras},
         year = 2013,
        month = nov,
       volume = {435},
       number = {4},
        pages = {2835-2860},
          doi = {10.1093/mnras/stt1458},
archivePrefix = {arXiv},
       eprint = {1308.3496},
 primaryClass = {astro-ph.CO},
       adsurl = {https://ui.adsabs.harvard.edu/abs/2013MNRAS.435.2835W}
}

@ARTICLE{Sampaio2022,
       author = {{Sampaio}, V.~M. and {de Carvalho}, R.~R. and {Ferreras}, I. and {Arag{\'o}n-Salamanca}, A. and {Parker}, L.~C.},
        title = "{From blue cloud to red sequence: evidence of morphological transition prior to star formation quenching}",
      journal = {\mnras},
         year = 2022,
        month = jan,
       volume = {509},
       number = {1},
        pages = {567-585},
          doi = {10.1093/mnras/stab3018},
archivePrefix = {arXiv},
       eprint = {2110.04342},
 primaryClass = {astro-ph.GA},
       adsurl = {https://ui.adsabs.harvard.edu/abs/2022MNRAS.509..567S}
}

@ARTICLE{Sampaio2023,
       author = {{Sampaio}, V.~M. and {Arag{\'o}n-Salamanca}, A. and {Merrifield}, M.~R. and {de Carvalho}, R.~R. and {Zhou}, S. and {Ferreras}, I.},
        title = "{The co-evolution of strong AGN and central galaxies in different environments}",
      journal = {\mnras},
         year = 2023,
        month = oct,
       volume = {524},
       number = {4},
        pages = {5327-5339},
          doi = {10.1093/mnras/stad2211},
archivePrefix = {arXiv},
       eprint = {2307.10435},
 primaryClass = {astro-ph.GA},
       adsurl = {https://ui.adsabs.harvard.edu/abs/2023MNRAS.524.5327S}
}

@ARTICLE{CidFernandes2010,
       author = {{Cid Fernandes}, R. and {Stasi{\'n}ska}, G. and {Schlickmann}, M.~S. and {Mateus}, A. and {Vale Asari}, N. and {Schoenell}, W. and {Sodr{\'e}}, L.},
        title = "{Alternative diagnostic diagrams and the `forgotten' population of weak line galaxies in the SDSS}",
      journal = {\mnras},
         year = 2010,
        month = apr,
       volume = {403},
       number = {2},
        pages = {1036-1053},
          doi = {10.1111/j.1365-2966.2009.16185.x},
archivePrefix = {arXiv},
       eprint = {0912.1643},
 primaryClass = {astro-ph.CO},
       adsurl = {https://ui.adsabs.harvard.edu/abs/2010MNRAS.403.1036C}
}

@ARTICLE{Baldwin1981,
       author = {{Baldwin}, J.~A. and {Phillips}, M.~M. and {Terlevich}, R.},
        title = "{Classification parameters for the emission-line spectra of extragalactic objects.}",
      journal = {\pasp},
         year = 1981,
        month = feb,
       volume = {93},
        pages = {5-19},
          doi = {10.1086/130766},
       adsurl = {https://ui.adsabs.harvard.edu/abs/1981PASP...93....5B}
}

@ARTICLE{Gunn1972,
   author = {{Gunn}, J.~E. and {Gott}, III, J.~R.},
    title = "{On the Infall of Matter Into Clusters of Galaxies and Some Effects on Their Evolution}",
  journal = {\apj},
     year = 1972,
    month = aug,
   volume = 176,
    pages = {1},
      doi = {10.1086/151605},
   adsurl = {http://adsabs.harvard.edu/abs/1972ApJ...176....1G}
}

@ARTICLE{Poggianti2016,
   author = {{Poggianti}, B.~M. and {Fasano}, G. and {Omizzolo}, A. and {Gullieuszik}, M. and 
	{Bettoni}, D. and {Moretti}, A. and {Paccagnella}, A. and {Jaff{\'e}}, Y.~L. and 
	{Vulcani}, B. and {Fritz}, J. and {Couch}, W. and {D'Onofrio}, M.
	},
    title = "{Jellyfish Galaxy Candidates at Low Redshift}",
  journal = {\aj},
archivePrefix = "arXiv",
   eprint = {1504.07105},
     year = 2016,
    month = mar,
   volume = 151,
      eid = {78},
    pages = {78},
      doi = {10.3847/0004-6256/151/3/78},
   adsurl = {http://adsabs.harvard.edu/abs/2016AJ....151...78P}
}

@ARTICLE{Fossati2016,
   author = {{Fossati}, M. and {Fumagalli}, M. and {Boselli}, A. and {Gavazzi}, G. and 
	{Sun}, M. and {Wilman}, D.~J.},
    title = "{MUSE sneaks a peek at extreme ram-pressure stripping events - II. The physical properties of the gas tail of ESO137-001}",
  journal = {\mnras},
archivePrefix = "arXiv",
   eprint = {1510.04283},
     year = 2016,
    month = jan,
   volume = 455,
    pages = {2028-2041},
      doi = {10.1093/mnras/stv2400},
   adsurl = {http://adsabs.harvard.edu/abs/2016MNRAS.455.2028F}
}

@ARTICLE{Jaffe2015,
   author = {{Jaff{\'e}}, Y.~L. and {Smith}, R. and {Candlish}, G.~N. and 
	{Poggianti}, B.~M. and {Sheen}, Y.-K. and {Verheijen}, M.~A.~W.
	},
    title = "{BUDHIES II: a phase-space view of H I gas stripping and star formation quenching in cluster galaxies}",
  journal = {\mnras},
archivePrefix = "arXiv",
   eprint = {1501.03819},
     year = 2015,
    month = apr,
   volume = 448,
    pages = {1715-1728},
      doi = {10.1093/mnras/stv100},
   adsurl = {http://adsabs.harvard.edu/abs/2015MNRAS.448.1715J}
}

@ARTICLE{Moore1996,
   author = {{Moore}, B. and {Katz}, N. and {Lake}, G. and {Dressler}, A. and 
	{Oemler}, A.},
    title = "{Galaxy harassment and the evolution of clusters of galaxies}",
  journal = {\nat},
   eprint = {astro-ph/9510034},
     year = 1996,
    month = feb,
   volume = 379,
    pages = {613-616},
      doi = {10.1038/379613a0},
   adsurl = {http://adsabs.harvard.edu/abs/1996Natur.379..613M}
}

@ARTICLE{NFW,
   author = {{Navarro}, J.~F. and {Frenk}, C.~S. and {White}, S.~D.~M.},
    title = "{The Structure of Cold Dark Matter Halos}",
  journal = {\apj},
   eprint = {astro-ph/9508025},
     year = 1996,
    month = may,
   volume = 462,
    pages = {563},
      doi = {10.1086/177173},
   adsurl = {http://adsabs.harvard.edu/abs/1996ApJ...462..563N}
}

@ARTICLE{Vollmer2001,
   author = {{Vollmer}, B. and {Cayatte}, V. and {Balkowski}, C. and {Duschl}, W.~J.
	},
    title = "{Ram Pressure Stripping and Galaxy Orbits: The Case of the Virgo Cluster}",
  journal = {\apj},
   eprint = {astro-ph/0107237},
     year = 2001,
    month = nov,
   volume = 561,
    pages = {708-726},
      doi = {10.1086/323368},
   adsurl = {http://adsabs.harvard.edu/abs/2001ApJ...561..708V}
}

@ARTICLE{Kenney2004,
   author = {{Kenney}, J.~D.~P. and {van Gorkom}, J.~H. and {Vollmer}, B.
	},
    title = "{VLA H I Observations of Gas Stripping in the Virgo Cluster Spiral NGC 4522}",
  journal = {\aj},
   eprint = {astro-ph/0403103},
     year = 2004,
    month = jun,
   volume = 127,
    pages = {3361-3374},
      doi = {10.1086/420805},
   adsurl = {http://adsabs.harvard.edu/abs/2004AJ....127.3361K}
}

@ARTICLE{Spitzer1951,
   author = {{Spitzer}, Jr., L. and {Baade}, W.},
    title = "{Stellar Populations and Collisions of Galaxies.}",
  journal = {\apj},
     year = 1951,
    month = mar,
   volume = 113,
    pages = {413},
      doi = {10.1086/145406},
   adsurl = {http://adsabs.harvard.edu/abs/1951ApJ...113..413S}
}

@ARTICLE{Merritt1983,
   author = {{Merritt}, D.},
    title = "{Relaxation and tidal stripping in rich clusters of galaxies. I. Evolution of the mass distribution}",
  journal = {\apj},
     year = 1983,
    month = jan,
   volume = 264,
    pages = {24-48},
      doi = {10.1086/160571},
   adsurl = {http://adsabs.harvard.edu/abs/1983ApJ...264...24M}
}

@ARTICLE{Valluri1993,
   author = {{Valluri}, M.},
    title = "{Compressive tidal heating of a disk galaxy in a rich cluster}",
  journal = {\apj},
     year = 1993,
    month = may,
   volume = 408,
    pages = {57-70},
      doi = {10.1086/172569},
   adsurl = {http://adsabs.harvard.edu/abs/1993ApJ...408...57V}
}

@ARTICLE{Boselli2006,
   author = {{Boselli}, A. and {Gavazzi}, G.},
    title = "{Environmental Effects on Late-Type Galaxies in Nearby Clusters}",
  journal = {\pasp},
   eprint = {astro-ph/0601108},
     year = 2006,
    month = apr,
   volume = 118,
    pages = {517-559},
      doi = {10.1086/500691},
   adsurl = {http://adsabs.harvard.edu/abs/2006PASP..118..517B}
}

@ARTICLE{Gullieuszik2015,
   author = {{Gullieuszik}, M. and {Poggianti}, B. and {Fasano}, G. and {Zaggia}, S. and 
	{Paccagnella}, A. and {Moretti}, A. and {Bettoni}, D. and {D'Onofrio}, M. and 
	{Couch}, W.~J. and {Vulcani}, B. and {Fritz}, J. and {Omizzolo}, A. and 
	{Baruffolo}, A. and {Schipani}, P. and {Capaccioli}, M. and 
	{Varela}, J.},
    title = "{OmegaWINGS: OmegaCAM-VST observations of WINGS galaxy clusters}",
  journal = {\aap},
archivePrefix = "arXiv",
   eprint = {1503.02628},
     year = 2015,
    month = sep,
   volume = 581,
      eid = {A41},
    pages = {A41},
      doi = {10.1051/0004-6361/201526061},
   adsurl = {http://adsabs.harvard.edu/abs/2015A%26A...581A..41G}
}

@ARTICLE{Fasano2006,
   author = {{Fasano}, G. and {Marmo}, C. and {Varela}, J. and {D'Onofrio}, M. and 
	{Poggianti}, B.~M. and {Moles}, M. and {Pignatelli}, E. and 
	{Bettoni}, D. and {Kj{\ae}rgaard}, P. and {Rizzi}, L. and {Couch}, W.~J. and 
	{Dressler}, A.},
    title = "{WINGS: a WIde-field Nearby Galaxy-cluster Survey. I. Optical imaging}",
  journal = {\aap},
   eprint = {astro-ph/0507247},
     year = 2006,
    month = jan,
   volume = 445,
    pages = {805-817},
      doi = {10.1051/0004-6361:20053816},
   adsurl = {http://adsabs.harvard.edu/abs/2006A%26A...445..805F}
}

@ARTICLE{McPartland2016,
   author = {{McPartland}, C. and {Ebeling}, H. and {Roediger}, E. and {Blumenthal}, K.
	},
    title = "{Jellyfish: the origin and distribution of extreme ram-pressure stripping events in massive galaxy clusters}",
  journal = {\mnras},
archivePrefix = "arXiv",
   eprint = {1511.00033},
     year = 2016,
    month = jan,
   volume = 455,
    pages = {2994-3008},
      doi = {10.1093/mnras/stv2508},
   adsurl = {http://adsabs.harvard.edu/abs/2016MNRAS.455.2994M}
}

@INPROCEEDINGS{Toomre1977,
   author = {{Toomre}, A.},
    title = "{Mergers and Some Consequences}",
booktitle = {Evolution of Galaxies and Stellar Populations},
     year = 1977,
publisher = {Yale University Observatory},
   editor = {{Tinsley}, B.~M. and {Larson}, D.~Campbell, R.~B.~G.},
    pages = {401},
   adsurl = {http://adsabs.harvard.edu/abs/1977egsp.conf..401T}
}

@ARTICLE{Tinsley1979,
   author = {{Tinsley}, B.~M. and {Larson}, R.~B.},
    title = "{Stellar population explosions in proto-elliptical galaxies}",
  journal = {\mnras},
     year = 1979,
    month = feb,
   volume = 186,
    pages = {503-517},
      doi = {10.1093/mnras/186.3.503},
   adsurl = {http://adsabs.harvard.edu/abs/1979MNRAS.186..503T}
}

@ARTICLE{Springel2000,
   author = {{Springel}, V.},
    title = "{Modelling star formation and feedback in simulations of interacting galaxies}",
  journal = {\mnras},
     year = 2000,
    month = mar,
   volume = 312,
    pages = {859-879},
      doi = {10.1046/j.1365-8711.2000.03187.x},
   adsurl = {http://adsabs.harvard.edu/abs/2000MNRAS.312..859S}
}

@ARTICLE{Byrd1990,
   author = {{Byrd}, G. and {Valtonen}, M.},
    title = "{Tidal generation of active spirals and S0 galaxies by rich clusters}",
  journal = {\apj},
     year = 1990,
    month = feb,
   volume = 350,
    pages = {89-94},
      doi = {10.1086/168362},
   adsurl = {http://adsabs.harvard.edu/abs/1990ApJ...350...89B}
}

@ARTICLE{Poggianti2017,
   author = {{Poggianti}, B.~M. and {Moretti}, A. and {Gullieuszik}, M. and 
	{Fritz}, J. and {Jaff{\'e}}, Y. and {Bettoni}, D. and {Fasano}, G. and 
	{Bellhouse}, C. and {Hau}, G. and {Vulcani}, B. and {Biviano}, A. and 
	{Omizzolo}, A. and {Paccagnella}, A. and {D'Onofrio}, M. and 
	{Cava}, A. and {Sheen}, Y.-K. and {Couch}, W. and {Owers}, M.
	},
    title = "{GASP. I. Gas Stripping Phenomena in Galaxies with MUSE}",
  journal = {\apj},
archivePrefix = "arXiv",
   eprint = {1704.05086},
     year = 2017,
    month = jul,
   volume = 844,
      eid = {48},
    pages = {48},
      doi = {10.3847/1538-4357/aa78ed},
   adsurl = {http://adsabs.harvard.edu/abs/2017ApJ...844...48P}
}

@ARTICLE{Poggianti2017b,
   author = {{Poggianti}, B.~M. and {Jaff{\'e}}, Y.~L. and {Moretti}, A. and 
	{Gullieuszik}, M. and {Radovich}, M. and {Tonnesen}, S. and 
	{Fritz}, J. and {Bettoni}, D. and {Vulcani}, B. and {Fasano}, G. and 
	{Bellhouse}, C. and {Hau}, G. and {Omizzolo}, A.},
    title = "{Ram-pressure feeding of supermassive black holes}",
  journal = {\nat},
archivePrefix = "arXiv",
   eprint = {1708.09036},
     year = 2017,
    month = aug,
   volume = 548,
    pages = {304-309},
      doi = {10.1038/nature23462},
   adsurl = {http://adsabs.harvard.edu/abs/2017Natur.548..304P}
}

@ARTICLE{Jaffe2018,
   author = {{Jaff{\'e}}, Y.~L. and {Poggianti}, B.~M. and {Moretti}, A. and 
	{Gullieuszik}, M. and {Smith}, R. and {Vulcani}, B. and {Fasano}, G. and 
	{Fritz}, J. and {Tonnesen}, S. and {Bettoni}, D. and {Hau}, G. and 
	{Biviano}, A. and {Bellhouse}, C. and {McGee}, S.},
    title = "{GASP. IX. Jellyfish galaxies in phase-space: an orbital study of intense ram-pressure stripping in clusters}",
  journal = {\mnras},
archivePrefix = "arXiv",
   eprint = {1802.07297},
     year = 2018,
    month = jun,
   volume = 476,
    pages = {4753-4764},
      doi = {10.1093/mnras/sty500},
   adsurl = {http://adsabs.harvard.edu/abs/2018MNRAS.476.4753J}
}

@ARTICLE{Vulcani2018,
   author = {{Vulcani}, B. and {Poggianti}, B.~M. and {Gullieuszik}, M. and 
  {Moretti}, A. and {Tonnesen}, S. and {Jaff{\'e}}, Y.~L. and 
  {Fritz}, J. and {Fasano}, G. and {Bettoni}, D.},
    title = "{Enhanced Star Formation in Both Disks and Ram-pressure-stripped Tails of GASP Jellyfish Galaxies}",
  journal = {\apjl},
archivePrefix = "arXiv",
   eprint = {1810.05164},
     year = 2018,
    month = oct,
   volume = 866,
      eid = {L25},
    pages = {L25},
      doi = {10.3847/2041-8213/aae68b},
   adsurl = {http://adsabs.harvard.edu/abs/2018ApJ...866L..25V}
}

@ARTICLE{Roediger2014,
   author = {{Roediger}, E. and {Br{\"u}ggen}, M. and {Owers}, M.~S. and 
	{Ebeling}, H. and {Sun}, M.},
    title = "{Star formation in shocked cluster spirals and their tails}",
  journal = {\mnras},
archivePrefix = "arXiv",
   eprint = {1405.1033},
     year = 2014,
    month = sep,
   volume = 443,
    pages = {L114-L118},
      doi = {10.1093/mnrasl/slu087},
   adsurl = {http://adsabs.harvard.edu/abs/2014MNRAS.443L.114R}
}

@ARTICLE{Mihos1993,
       author = {{Mihos}, J. Christopher and {Bothun}, Gregory D. and
         {Richstone}, Douglas O.},
        title = "{Modeling the Spatial Distribution of Star Formation in Interacting Disk Galaxies}",
      journal = {\apj},
         year = "1993",
        month = "Nov",
       volume = {418},
        pages = {82},
          doi = {10.1086/173373},
       adsurl = {https://ui.adsabs.harvard.edu/abs/1993ApJ...418...82M}
}

@ARTICLE{Poggianti2019b,
       author = {{Poggianti}, Bianca M. and {Ignesti}, Alessandro and {Gitti}, Myriam and
         {Wolter}, Anna and {Brighenti}, Fabrizio and {Biviano}, Andrea and
         {George}, Koshy and {Vulcani}, Benedetta and {Gullieuszik}, Marco and
         {Moretti}, Alessia and {Paladino}, Rosita and {Bettoni}, Daniela and
         {Franchetto}, Andrea and {Jaff{\'e}}, Yara L. and {Radovich}, Mario and
         {Roediger}, Elke and {Tomi{\v{c}}i{\'c}}, Neven and
         {Tonnesen}, Stephanie and {Bellhouse}, Callum and {Fritz}, Jacopo and
         {Omizzolo}, Alessandro},
        title = "{GASP XXIII: A Jellyfish Galaxy as an Astrophysical Laboratory of the Baryonic Cycle}",
      journal = {\apj},
         year = 2019,
        month = dec,
       volume = {887},
       number = {2},
          eid = {155},
        pages = {155},
          doi = {10.3847/1538-4357/ab5224},
archivePrefix = {arXiv},
       eprint = {1910.11622},
 primaryClass = {astro-ph.GA},
       adsurl = {https://ui.adsabs.harvard.edu/abs/2019ApJ...887..155P}
}

@ARTICLE{Tomicic2018,
       author = {{Tomi{\v{c}}i{\'c}}, Neven and {Hughes}, Annie and {Kreckel}, Kathryn and
         {Renaud}, Florent and {Pety}, J{\'e}r{\^o}me and {Schinnerer}, Eva and
         {Saito}, Toshiki and {Querejeta}, Miguel and {Faesi}, Christopher M. and
         {Garcia-Burillo}, Santiago},
        title = "{Two Orders of Magnitude Variation in the Star Formation Efficiency across the Premerger Galaxy NGC 2276}",
      journal = {\apjl},
         year = 2018,
        month = dec,
       volume = {869},
       number = {2},
          eid = {L38},
        pages = {L38},
          doi = {10.3847/2041-8213/aaf810},
archivePrefix = {arXiv},
       eprint = {1812.05048},
 primaryClass = {astro-ph.GA},
       adsurl = {https://ui.adsabs.harvard.edu/abs/2018ApJ...869L..38T}
}

@ARTICLE{Vulcani2020,
       author = {{Vulcani}, Benedetta and {Fritz}, Jacopo and {Poggianti}, Bianca M. and {Bettoni}, Daniela and {Franchetto}, Andrea and {Moretti}, Alessia and {Gullieuszik}, Marco and {Jaff{\'e}}, Yara and {Biviano}, Andrea and {Radovich}, Mario and {Mingozzi}, Matilde},
        title = "{GASP XXIV. The History of Abruptly Quenched Galaxies in Clusters}",
      journal = {\apj},
         year = 2020,
        month = apr,
       volume = {892},
       number = {2},
          eid = {146},
        pages = {146},
          doi = {10.3847/1538-4357/ab7bdd},
archivePrefix = {arXiv},
       eprint = {2003.02274},
 primaryClass = {astro-ph.GA},
       adsurl = {https://ui.adsabs.harvard.edu/abs/2020ApJ...892..146V}
}

@ARTICLE{Sheen2017,
       author = {{Sheen}, Yun-Kyeong and {Smith}, Rory and {Jaff{\'e}}, Yara and
         {Kim}, Minjin and {Yi}, Sukyoung K. and {Duc}, Pierre-Alain and
         {Nantais}, Julie and {Candlish}, Graeme and {Demarco}, Ricardo and
         {Treister}, Ezequiel},
        title = "{Discovery of Ram-pressure Stripped Gas around an Elliptical Galaxy in Abell 2670}",
      journal = {\apjl},
         year = 2017,
        month = may,
       volume = {840},
       number = {1},
          eid = {L7},
        pages = {L7},
          doi = {10.3847/2041-8213/aa6d79},
archivePrefix = {arXiv},
       eprint = {1704.05173},
 primaryClass = {astro-ph.GA},
       adsurl = {https://ui.adsabs.harvard.edu/abs/2017ApJ...840L...7S}
}

@ARTICLE{Roberts2020,
       author = {{Roberts}, Ian D. and {Parker}, Laura C.},
        title = "{Ram pressure stripping candidates in the coma cluster: evidence for enhanced star formation}",
      journal = {\mnras},
         year = 2020,
        month = jun,
       volume = {495},
       number = {1},
        pages = {554-569},
          doi = {10.1093/mnras/staa1213},
archivePrefix = {arXiv},
       eprint = {2004.12033},
 primaryClass = {astro-ph.GA},
       adsurl = {https://ui.adsabs.harvard.edu/abs/2020MNRAS.495..554R}
}

@ARTICLE{Bellhouse2021,
       author = {{Bellhouse}, Callum and {McGee}, Sean L. and {Smith}, Rory and {Poggianti}, Bianca M. and {Jaff{\'e}}, Yara L. and {Kraljic}, Katarina and {Franchetto}, Andrea and {Fritz}, Jacopo and {Vulcani}, Benedetta and {Tonnesen}, Stephanie and {Roediger}, Elke and {Moretti}, Alessia and {Gullieuszik}, Marco and {Shin}, Jihye},
        title = "{GASP XXIX - unwinding the arms of spiral galaxies via ram-pressure stripping}",
      journal = {\mnras},
         year = 2021,
        month = jan,
       volume = {500},
       number = {1},
        pages = {1285-1312},
          doi = {10.1093/mnras/staa3298},
archivePrefix = {arXiv},
       eprint = {2010.09733},
 primaryClass = {astro-ph.GA},
       adsurl = {https://ui.adsabs.harvard.edu/abs/2021MNRAS.500.1285B}
}

@ARTICLE{Stroe2020,
       author = {{Stroe}, Andra and {Hussaini}, Maryam and {Husemann}, Bernd and {Sobral}, David and {Tremblay}, Grant},
        title = "{The First Integral Field Unit Spectroscopic View of Shocked Cluster Galaxies}",
      journal = {\apjl},
         year = 2020,
        month = dec,
       volume = {905},
       number = {2},
          eid = {L22},
        pages = {L22},
          doi = {10.3847/2041-8213/abcb04},
archivePrefix = {arXiv},
       eprint = {2011.13935},
 primaryClass = {astro-ph.GA},
       adsurl = {https://ui.adsabs.harvard.edu/abs/2020ApJ...905L..22S}
}

@ARTICLE{Vulcani2022,
       author = {{Vulcani}, Benedetta and {Poggianti}, Bianca M. and {Smith}, Rory and {Moretti}, Alessia and {Jaff{\'e}}, Yara L. and {Gullieuszik}, Marco and {Fritz}, Jacopo and {Bellhouse}, Callum},
        title = "{The Relevance of Ram Pressure Stripping for the Evolution of Blue Cluster Galaxies as Seen at Optical Wavelengths}",
      journal = {\apj},
         year = 2022,
        month = mar,
       volume = {927},
       number = {1},
          eid = {91},
        pages = {91},
          doi = {10.3847/1538-4357/ac4809},
archivePrefix = {arXiv},
       eprint = {2201.02644},
 primaryClass = {astro-ph.GA},
       adsurl = {https://ui.adsabs.harvard.edu/abs/2022ApJ...927...91V}
}

@ARTICLE{Walmsley2022,
       author = {{Walmsley}, Mike and {Lintott}, Chris and {G{\'e}ron}, Tobias and {Kruk}, Sandor and {Krawczyk}, Coleman and {Willett}, Kyle W. and {Bamford}, Steven and {Kelvin}, Lee S. and {Fortson}, Lucy and {Gal}, Yarin and {Keel}, William and {Masters}, Karen L. and {Mehta}, Vihang and {Simmons}, Brooke D. and {Smethurst}, Rebecca and {Smith}, Lewis and {Baeten}, Elisabeth M. and {Macmillan}, Christine},
        title = "{Galaxy Zoo DECaLS: Detailed visual morphology measurements from volunteers and deep learning for 314 000 galaxies}",
      journal = {\mnras},
         year = 2022,
        month = jan,
       volume = {509},
       number = {3},
        pages = {3966-3988},
          doi = {10.1093/mnras/stab2093},
archivePrefix = {arXiv},
       eprint = {2102.08414},
 primaryClass = {astro-ph.GA},
       adsurl = {https://ui.adsabs.harvard.edu/abs/2022MNRAS.509.3966W}
}

@ARTICLE{Lintott2008,
       author = {{Lintott}, Chris J. and {Schawinski}, Kevin and {Slosar}, An{\v{z}}e and {Land}, Kate and {Bamford}, Steven and {Thomas}, Daniel and {Raddick}, M. Jordan and {Nichol}, Robert C. and {Szalay}, Alex and {Andreescu}, Dan and {Murray}, Phil and {Vandenberg}, Jan},
        title = "{Galaxy Zoo: morphologies derived from visual inspection of galaxies from the Sloan Digital Sky Survey}",
      journal = {\mnras},
         year = 2008,
        month = sep,
       volume = {389},
       number = {3},
        pages = {1179-1189},
          doi = {10.1111/j.1365-2966.2008.13689.x},
archivePrefix = {arXiv},
       eprint = {0804.4483},
 primaryClass = {astro-ph},
       adsurl = {https://ui.adsabs.harvard.edu/abs/2008MNRAS.389.1179L}
}

@ARTICLE{Dey2019,
       author = {{Dey}, Arjun and {Schlegel}, David J. and {Lang}, Dustin and {Blum}, Robert and {Burleigh}, Kaylan and {Fan}, Xiaohui and {Findlay}, Joseph R. and {Finkbeiner}, Doug and {Herrera}, David and {Juneau}, St{\'e}phanie and {Landriau}, Martin and {Levi}, Michael and {McGreer}, Ian and {Meisner}, Aaron and {Myers}, Adam D. and {Moustakas}, John and {Nugent}, Peter and {Patej}, Anna and {Schlafly}, Edward F. and {Walker}, Alistair R. and {Valdes}, Francisco and {Weaver}, Benjamin A. and {Y{\`e}che}, Christophe and {Zou}, Hu and {Zhou}, Xu and {Abareshi}, Behzad and {Abbott}, T.~M.~C. and {Abolfathi}, Bela and {Aguilera}, C. and {Alam}, Shadab and {Allen}, Lori and {Alvarez}, A. and {Annis}, James and {Ansarinejad}, Behzad and {Aubert}, Marie and {Beechert}, Jacqueline and {Bell}, Eric F. and {BenZvi}, Segev Y. and {Beutler}, Florian and {Bielby}, Richard M. and {Bolton}, Adam S. and {Brice{\~n}o}, C{\'e}sar and {Buckley-Geer}, Elizabeth J. and {Butler}, Karen and {Calamida}, Annalisa and {Carlberg}, Raymond G. and {Carter}, Paul and {Casas}, Ricard and {Castander}, Francisco J. and {Choi}, Yumi and {Comparat}, Johan and {Cukanovaite}, Elena and {Delubac}, Timoth{\'e}e and {DeVries}, Kaitlin and {Dey}, Sharmila and {Dhungana}, Govinda and {Dickinson}, Mark and {Ding}, Zhejie and {Donaldson}, John B. and {Duan}, Yutong and {Duckworth}, Christopher J. and {Eftekharzadeh}, Sarah and {Eisenstein}, Daniel J. and {Etourneau}, Thomas and {Fagrelius}, Parker A. and {Farihi}, Jay and {Fitzpatrick}, Mike and {Font-Ribera}, Andreu and {Fulmer}, Leah and {G{\"a}nsicke}, Boris T. and {Gaztanaga}, Enrique and {George}, Koshy and {Gerdes}, David W. and {Gontcho}, Satya Gontcho A. and {Gorgoni}, Claudio and {Green}, Gregory and {Guy}, Julien and {Harmer}, Diane and {Hernandez}, M. and {Honscheid}, Klaus and {Huang}, Lijuan Wendy and {James}, David J. and {Jannuzi}, Buell T. and {Jiang}, Linhua and {Joyce}, Richard and {Karcher}, Armin and {Karkar}, Sonia and {Kehoe}, Robert and {Kneib}, Jean-Paul and {Kueter-Young}, Andrea and {Lan}, Ting-Wen and {Lauer}, Tod R. and {Le Guillou}, Laurent and {Le Van Suu}, Auguste and {Lee}, Jae Hyeon and {Lesser}, Michael and {Perreault Levasseur}, Laurence and {Li}, Ting S. and {Mann}, Justin L. and {Marshall}, Robert and {Mart{\'\i}nez-V{\'a}zquez}, C.~E. and {Martini}, Paul and {du Mas des Bourboux}, H{\'e}lion and {McManus}, Sean and {Meier}, Tobias Gabriel and {M{\'e}nard}, Brice and {Metcalfe}, Nigel and {Mu{\~n}oz-Guti{\'e}rrez}, Andrea and {Najita}, Joan and {Napier}, Kevin and {Narayan}, Gautham and {Newman}, Jeffrey A. and {Nie}, Jundan and {Nord}, Brian and {Norman}, Dara J. and {Olsen}, Knut A.~G. and {Paat}, Anthony and {Palanque-Delabrouille}, Nathalie and {Peng}, Xiyan and {Poppett}, Claire L. and {Poremba}, Megan R. and {Prakash}, Abhishek and {Rabinowitz}, David and {Raichoor}, Anand and {Rezaie}, Mehdi and {Robertson}, A.~N. and {Roe}, Natalie A. and {Ross}, Ashley J. and {Ross}, Nicholas P. and {Rudnick}, Gregory and {Safonova}, Sasha and {Saha}, Abhijit and {S{\'a}nchez}, F. Javier and {Savary}, Elodie and {Schweiker}, Heidi and {Scott}, Adam and {Seo}, Hee-Jong and {Shan}, Huanyuan and {Silva}, David R. and {Slepian}, Zachary and {Soto}, Christian and {Sprayberry}, David and {Staten}, Ryan and {Stillman}, Coley M. and {Stupak}, Robert J. and {Summers}, David L. and {Sien Tie}, Suk and {Tirado}, H. and {Vargas-Maga{\~n}a}, Mariana and {Vivas}, A. Katherina and {Wechsler}, Risa H. and {Williams}, Doug and {Yang}, Jinyi and {Yang}, Qian and {Yapici}, Tolga and {Zaritsky}, Dennis and {Zenteno}, A. and {Zhang}, Kai and {Zhang}, Tianmeng and {Zhou}, Rongpu and {Zhou}, Zhimin},
        title = "{Overview of the DESI Legacy Imaging Surveys}",
      journal = {\aj},
         year = 2019,
        month = may,
       volume = {157},
       number = {5},
          eid = {168},
        pages = {168},
          doi = {10.3847/1538-3881/ab089d},
archivePrefix = {arXiv},
       eprint = {1804.08657},
 primaryClass = {astro-ph.IM},
       adsurl = {https://ui.adsabs.harvard.edu/abs/2019AJ....157..168D}
}

@ARTICLE{Kolcu2022,
       author = {{Kolcu}, Tutku and {Crossett}, Jacob P. and {Bellhouse}, Callum and {McGee}, Sean},
        title = "{Quantifying the role of ram-pressure stripping of galaxies within galaxy groups}",
      journal = {\mnras},
         year = 2022,
        month = oct,
       volume = {515},
       number = {4},
        pages = {5877-5893},
          doi = {10.1093/mnras/stac2177},
archivePrefix = {arXiv},
       eprint = {2208.01666},
 primaryClass = {astro-ph.GA},
       adsurl = {https://ui.adsabs.harvard.edu/abs/2022MNRAS.515.5877K}
}

@ARTICLE{Finoguenov2020,
       author = {{Finoguenov}, A. and {Rykoff}, E. and {Clerc}, N. and {Costanzi}, M. and {Hagstotz}, S. and {Ider Chitham}, J. and {Kiiveri}, K. and {Kirkpatrick}, C.~C. and {Capasso}, R. and {Comparat}, J. and {Damsted}, S. and {Dupke}, R. and {Erfanianfar}, G. and {Patrick Henry}, J. and {Kaefer}, F. and {Kneib}, J. -P. and {Lindholm}, V. and {Rozo}, E. and {van Waerbeke}, L. and {Weller}, J.},
        title = "{CODEX clusters. Survey, catalog, and cosmology of the X-ray luminosity function}",
      journal = {\aap},
         year = 2020,
        month = jun,
       volume = {638},
          eid = {A114},
        pages = {A114},
          doi = {10.1051/0004-6361/201937283},
archivePrefix = {arXiv},
       eprint = {1912.03262},
 primaryClass = {astro-ph.CO},
       adsurl = {https://ui.adsabs.harvard.edu/abs/2020A&A...638A.114F}
}

@ARTICLE{Piffaretti2011,
       author = {{Piffaretti}, R. and {Arnaud}, M. and {Pratt}, G.~W. and {Pointecouteau}, E. and {Melin}, J. -B.},
        title = "{The MCXC: a meta-catalogue of x-ray detected clusters of galaxies}",
      journal = {\aap},
         year = 2011,
        month = oct,
       volume = {534},
          eid = {A109},
        pages = {A109},
          doi = {10.1051/0004-6361/201015377},
archivePrefix = {arXiv},
       eprint = {1007.1916},
 primaryClass = {astro-ph.CO},
       adsurl = {https://ui.adsabs.harvard.edu/abs/2011A&A...534A.109P}
}

@ARTICLE{Crossett2025,
       author = {{Crossett}, Jacob P. and {Jaff{\'e}}, Yara L. and {McGee}, Sean L. and {Smith}, Rory and {Bellhouse}, Callum and {Bettoni}, Daniela and {Vulcani}, Benedetta and {Kelkar}, Kshitija and {Louren{\c{c}}o}, Ana C.~C.},
        title = "{Identification of ram pressure stripping features in galaxies using citizen science}",
      journal = {\aap},
         year = 2025,
        month = feb,
       volume = {694},
          eid = {A204},
        pages = {A204},
          doi = {10.1051/0004-6361/202450371},
archivePrefix = {arXiv},
       eprint = {2412.10060},
 primaryClass = {astro-ph.GA},
       adsurl = {https://ui.adsabs.harvard.edu/abs/2025A&A...694A.204C}
}

@ARTICLE{Zinger2024,
       author = {{Zinger}, Elad and {Joshi}, Gandhali D. and {Pillepich}, Annalisa and {Rohr}, Eric and {Nelson}, Dylan},
        title = "{Jellyfish galaxies with the IllustrisTNG simulations - citizen-science results towards large distances, low-mass hosts, and high redshifts}",
      journal = {\mnras},
         year = 2024,
        month = jan,
       volume = {527},
       number = {3},
        pages = {8257-8289},
          doi = {10.1093/mnras/stad3716},
archivePrefix = {arXiv},
       eprint = {2304.09202},
 primaryClass = {astro-ph.GA},
       adsurl = {https://ui.adsabs.harvard.edu/abs/2024MNRAS.527.8257Z}
}

@ARTICLE{York2000,
       author = {{York}, Donald G. and {Adelman}, J. and {Anderson}, Jr., John E. and {Anderson}, Scott F. and {Annis}, James and {Bahcall}, Neta A. and {Bakken}, J.~A. and {Barkhouser}, Robert and {Bastian}, Steven and {Berman}, Eileen and {Boroski}, William N. and {Bracker}, Steve and {Briegel}, Charlie and {Briggs}, John W. and {Brinkmann}, J. and {Brunner}, Robert and {Burles}, Scott and {Carey}, Larry and {Carr}, Michael A. and {Castander}, Francisco J. and {Chen}, Bing and {Colestock}, Patrick L. and {Connolly}, A.~J. and {Crocker}, J.~H. and {Csabai}, Istv{\'a}n and {Czarapata}, Paul C. and {Davis}, John Eric and {Doi}, Mamoru and {Dombeck}, Tom and {Eisenstein}, Daniel and {Ellman}, Nancy and {Elms}, Brian R. and {Evans}, Michael L. and {Fan}, Xiaohui and {Federwitz}, Glenn R. and {Fiscelli}, Larry and {Friedman}, Scott and {Frieman}, Joshua A. and {Fukugita}, Masataka and {Gillespie}, Bruce and {Gunn}, James E. and {Gurbani}, Vijay K. and {de Haas}, Ernst and {Haldeman}, Merle and {Harris}, Frederick H. and {Hayes}, J. and {Heckman}, Timothy M. and {Hennessy}, G.~S. and {Hindsley}, Robert B. and {Holm}, Scott and {Holmgren}, Donald J. and {Huang}, Chi-hao and {Hull}, Charles and {Husby}, Don and {Ichikawa}, Shin-Ichi and {Ichikawa}, Takashi and {Ivezi{\'c}}, {\v{Z}}eljko and {Kent}, Stephen and {Kim}, Rita S.~J. and {Kinney}, E. and {Klaene}, Mark and {Kleinman}, A.~N. and {Kleinman}, S. and {Knapp}, G.~R. and {Korienek}, John and {Kron}, Richard G. and {Kunszt}, Peter Z. and {Lamb}, D.~Q. and {Lee}, B. and {Leger}, R. French and {Limmongkol}, Siriluk and {Lindenmeyer}, Carl and {Long}, Daniel C. and {Loomis}, Craig and {Loveday}, Jon and {Lucinio}, Rich and {Lupton}, Robert H. and {MacKinnon}, Bryan and {Mannery}, Edward J. and {Mantsch}, P.~M. and {Margon}, Bruce and {McGehee}, Peregrine and {McKay}, Timothy A. and {Meiksin}, Avery and {Merelli}, Aronne and {Monet}, David G. and {Munn}, Jeffrey A. and {Narayanan}, Vijay K. and {Nash}, Thomas and {Neilsen}, Eric and {Neswold}, Rich and {Newberg}, Heidi Jo and {Nichol}, R.~C. and {Nicinski}, Tom and {Nonino}, Mario and {Okada}, Norio and {Okamura}, Sadanori and {Ostriker}, Jeremiah P. and {Owen}, Russell and {Pauls}, A. George and {Peoples}, John and {Peterson}, R.~L. and {Petravick}, Donald and {Pier}, Jeffrey R. and {Pope}, Adrian and {Pordes}, Ruth and {Prosapio}, Angela and {Rechenmacher}, Ron and {Quinn}, Thomas R. and {Richards}, Gordon T. and {Richmond}, Michael W. and {Rivetta}, Claudio H. and {Rockosi}, Constance M. and {Ruthmansdorfer}, Kurt and {Sandford}, Dale and {Schlegel}, David J. and {Schneider}, Donald P. and {Sekiguchi}, Maki and {Sergey}, Gary and {Shimasaku}, Kazuhiro and {Siegmund}, Walter A. and {Smee}, Stephen and {Smith}, J. Allyn and {Snedden}, S. and {Stone}, R. and {Stoughton}, Chris and {Strauss}, Michael A. and {Stubbs}, Christopher and {SubbaRao}, Mark and {Szalay}, Alexander S. and {Szapudi}, Istvan and {Szokoly}, Gyula P. and {Thakar}, Anirudda R. and {Tremonti}, Christy and {Tucker}, Douglas L. and {Uomoto}, Alan and {Vanden Berk}, Dan and {Vogeley}, Michael S. and {Waddell}, Patrick and {Wang}, Shu-i. and {Watanabe}, Masaru and {Weinberg}, David H. and {Yanny}, Brian and {Yasuda}, Naoki and {SDSS Collaboration}},
        title = "{The Sloan Digital Sky Survey: Technical Summary}",
      journal = {\aj},
         year = 2000,
        month = sep,
       volume = {120},
       number = {3},
        pages = {1579-1587},
          doi = {10.1086/301513},
archivePrefix = {arXiv},
       eprint = {astro-ph/0006396},
 primaryClass = {astro-ph},
       adsurl = {https://ui.adsabs.harvard.edu/abs/2000AJ....120.1579Y}
}

@ARTICLE{Flaugher2015,
       author = {{Flaugher}, B. and {Diehl}, H.~T. and {Honscheid}, K. and {Abbott}, T.~M.~C. and {Alvarez}, O. and {Angstadt}, R. and {Annis}, J.~T. and {Antonik}, M. and {Ballester}, O. and {Beaufore}, L. and {Bernstein}, G.~M. and {Bernstein}, R.~A. and {Bigelow}, B. and {Bonati}, M. and {Boprie}, D. and {Brooks}, D. and {Buckley-Geer}, E.~J. and {Campa}, J. and {Cardiel-Sas}, L. and {Castander}, F.~J. and {Castilla}, J. and {Cease}, H. and {Cela-Ruiz}, J.~M. and {Chappa}, S. and {Chi}, E. and {Cooper}, C. and {da Costa}, L.~N. and {Dede}, E. and {Derylo}, G. and {DePoy}, D.~L. and {de Vicente}, J. and {Doel}, P. and {Drlica-Wagner}, A. and {Eiting}, J. and {Elliott}, A.~E. and {Emes}, J. and {Estrada}, J. and {Fausti Neto}, A. and {Finley}, D.~A. and {Flores}, R. and {Frieman}, J. and {Gerdes}, D. and {Gladders}, M.~D. and {Gregory}, B. and {Gutierrez}, G.~R. and {Hao}, J. and {Holland}, S.~E. and {Holm}, S. and {Huffman}, D. and {Jackson}, C. and {James}, D.~J. and {Jonas}, M. and {Karcher}, A. and {Karliner}, I. and {Kent}, S. and {Kessler}, R. and {Kozlovsky}, M. and {Kron}, R.~G. and {Kubik}, D. and {Kuehn}, K. and {Kuhlmann}, S. and {Kuk}, K. and {Lahav}, O. and {Lathrop}, A. and {Lee}, J. and {Levi}, M.~E. and {Lewis}, P. and {Li}, T.~S. and {Mandrichenko}, I. and {Marshall}, J.~L. and {Martinez}, G. and {Merritt}, K.~W. and {Miquel}, R. and {Mu{\~n}oz}, F. and {Neilsen}, E.~H. and {Nichol}, R.~C. and {Nord}, B. and {Ogando}, R. and {Olsen}, J. and {Palaio}, N. and {Patton}, K. and {Peoples}, J. and {Plazas}, A.~A. and {Rauch}, J. and {Reil}, K. and {Rheault}, J. -P. and {Roe}, N.~A. and {Rogers}, H. and {Roodman}, A. and {Sanchez}, E. and {Scarpine}, V. and {Schindler}, R.~H. and {Schmidt}, R. and {Schmitt}, R. and {Schubnell}, M. and {Schultz}, K. and {Schurter}, P. and {Scott}, L. and {Serrano}, S. and {Shaw}, T.~M. and {Smith}, R.~C. and {Soares-Santos}, M. and {Stefanik}, A. and {Stuermer}, W. and {Suchyta}, E. and {Sypniewski}, A. and {Tarle}, G. and {Thaler}, J. and {Tighe}, R. and {Tran}, C. and {Tucker}, D. and {Walker}, A.~R. and {Wang}, G. and {Watson}, M. and {Weaverdyck}, C. and {Wester}, W. and {Woods}, R. and {Yanny}, B. and {DES Collaboration}},
        title = "{The Dark Energy Camera}",
      journal = {\aj},
         year = 2015,
        month = nov,
       volume = {150},
       number = {5},
          eid = {150},
        pages = {150},
          doi = {10.1088/0004-6256/150/5/150},
archivePrefix = {arXiv},
       eprint = {1504.02900},
 primaryClass = {astro-ph.IM},
       adsurl = {https://ui.adsabs.harvard.edu/abs/2015AJ....150..150F}
}

@ARTICLE{Sifon2025,
       author = {{Sif{\'o}n}, Crist{\'o}bal and {Finoguenov}, Alexis and {Haines}, Christopher P. and {Jaff{\'e}}, Yara and {Amrutha}, B.~M. and {Demarco}, Ricardo and {Lima}, E.~V.~R. and {Lima-Dias}, Ciria and {M{\'e}ndez-Hern{\'a}ndez}, Hugo and {Merluzzi}, Paola and {Monachesi}, Antonela and {Teixeira}, Gabriel S.~M. and {Tejos}, Nicolas and {Almeida-Fernandes}, F. and {Araya-Araya}, Pablo and {Argudo-Fern{\'a}ndez}, Maria and {Baier-Soto}, Ra{\'u}l and {Bilton}, Lawrence E. and {Bom}, C.~R. and {Calder{\'o}n}, Juan Pablo and {Cassar{\`a}}, Letizia P. and {Comparat}, Johan and {Courtois}, H.~M. and {D'Ago}, Giuseppe and {Dupuy}, Alexandra and {Fritz}, Alexander and {Haack}, Rodrigo F. and {Herpich}, Fabio R. and {Ibar}, E. and {Kuchner}, Ulrike and {Lacerna}, Ivan and {Lopes}, Amanda R. and {Lopez}, Sebastian and {L{\"o}sch}, Elismar and {McGee}, Sean and {Mendes de Oliveira}, C. and {Morelli}, Lorenzo and {Moretti}, Alessia and {Pallero}, Diego and {Piraino-Cerda}, Franco and {Pompei}, Emanuela and {Rescigno}, U. and {Smith Castelli}, Anal{\'\i}a V. and {Smith}, Rory and {Sodr{\'e}}, Jr., Laerte and {Tempel}, Elmo},
        title = "{CHANCES, the Chilean Cluster Galaxy Evolution Survey: Selection and initial characterisation of clusters and superclusters}",
      journal = {\aap},
         year = 2025,
        month = may,
       volume = {697},
          eid = {A92},
        pages = {A92},
          doi = {10.1051/0004-6361/202452710},
archivePrefix = {arXiv},
       eprint = {2411.13655},
 primaryClass = {astro-ph.GA},
       adsurl = {https://ui.adsabs.harvard.edu/abs/2025A&A...697A..92S}
}

@ARTICLE{Roberts2021,
       author = {{Roberts}, I.~D. and {van Weeren}, R.~J. and {McGee}, S.~L. and {Botteon}, A. and {Drabent}, A. and {Ignesti}, A. and {Rottgering}, H.~J.~A. and {Shimwell}, T.~W. and {Tasse}, C.},
        title = "{LoTSS jellyfish galaxies. I. Radio tails in low redshift clusters}",
      journal = {\aap},
         year = 2021,
        month = jun,
       volume = {650},
          eid = {A111},
        pages = {A111},
          doi = {10.1051/0004-6361/202140784},
archivePrefix = {arXiv},
       eprint = {2104.05383},
 primaryClass = {astro-ph.GA},
       adsurl = {https://ui.adsabs.harvard.edu/abs/2021A&A...650A.111R}
}

@ARTICLE{Roberts2021b,
       author = {{Roberts}, I.~D. and {van Weeren}, R.~J. and {McGee}, S.~L. and {Botteon}, A. and {Ignesti}, A. and {Rottgering}, H.~J.~A.},
        title = "{LoTSS jellyfish galaxies. II. Ram pressure stripping in groups versus clusters}",
      journal = {\aap},
         year = 2021,
        month = aug,
       volume = {652},
          eid = {A153},
        pages = {A153},
          doi = {10.1051/0004-6361/202141118},
archivePrefix = {arXiv},
       eprint = {2106.06315},
 primaryClass = {astro-ph.GA},
       adsurl = {https://ui.adsabs.harvard.edu/abs/2021A&A...652A.153R}
}

@ARTICLE{Rhee2017,
       author = {{Rhee}, Jinsu and {Smith}, Rory and {Choi}, Hoseung and {Yi}, Sukyoung K. and {Jaff{\'e}}, Yara and {Candlish}, Graeme and {S{\'a}nchez-J{\'a}nssen}, Ruben},
        title = "{Phase-space Analysis in the Group and Cluster Environment: Time Since Infall and Tidal Mass Loss}",
      journal = {\apj},
         year = 2017,
        month = jul,
       volume = {843},
       number = {2},
          eid = {128},
        pages = {128},
          doi = {10.3847/1538-4357/aa6d6c},
archivePrefix = {arXiv},
       eprint = {1704.04243},
 primaryClass = {astro-ph.GA},
       adsurl = {https://ui.adsabs.harvard.edu/abs/2017ApJ...843..128R}
}

@ARTICLE{Poggianti2025,
       author = {{Poggianti}, Bianca M. and {Vulcani}, Benedetta and {Tomicic}, Neven and {Moretti}, Alessia and {Gullieuszik}, Marco and {Bacchini}, Cecilia and {Fritz}, Jacopo and {George}, Koshy and {Gitti}, Myriam and {Ignesti}, Alessandro and {Jaff{\'e}}, Yara and {Lassen}, Augusto and {Marasco}, Antonino and {Radovich}, Mario and {Serra}, Paolo and {Smith}, Rory and {Tonnesen}, Stephanie and {Wolter}, Anna},
        title = "{The MUSE view of ram pressure stripped galaxies in clusters: The GASP sample}",
      journal = {\aap},
         year = 2025,
        month = jul,
       volume = {699},
          eid = {A357},
        pages = {A357},
          doi = {10.1051/0004-6361/202554200},
archivePrefix = {arXiv},
       eprint = {2505.21107},
 primaryClass = {astro-ph.GA},
       adsurl = {https://ui.adsabs.harvard.edu/abs/2025A&A...699A.357P}
}

@ARTICLE{Salinas2024,
       author = {{Salinas}, Vicente and {Jaff{\'e}}, Yara L. and {Smith}, Rory and {Shinn}, Jong-Ho and {Crossett}, Jacob P. and {Gullieuszik}, Marco and {Gonz{\'a}lez-Tor{\`a}}, Gemma and {Piraino-Cerda}, Franco and {Poggianti}, Bianca and {Vulcani}, Benedetta and {Biviano}, Andrea and {Louren{\c{c}}o}, Ana C.~C. and {Bilton}, Lawrence E. and {Kelkar}, Kshitija and {Calder{\'o}n-Castillo}, Paula},
        title = "{Constraining the duration of ram pressure stripping features in the optical from the direction of jellyfish galaxy tails}",
      journal = {\mnras},
         year = 2024,
        month = sep,
       volume = {533},
       number = {1},
        pages = {341-359},
          doi = {10.1093/mnras/stae1784},
archivePrefix = {arXiv},
       eprint = {2408.03396},
 primaryClass = {astro-ph.GA},
       adsurl = {https://ui.adsabs.harvard.edu/abs/2024MNRAS.533..341S}
}

@ARTICLE{Biviano2024,
       author = {{Biviano}, Andrea and {Poggianti}, Bianca M. and {Jaff{\'e}}, Yara and {Louren{\c{c}}o}, Ana C. and {Pizzuti}, Lorenzo and {Moretti}, Alessia and {Vulcani}, Benedetta},
        title = "{The Radial Orbits of Ram-pressure-stripped Galaxies in Clusters from the GASP Survey}",
      journal = {\apj},
         year = 2024,
        month = apr,
       volume = {965},
       number = {2},
          eid = {117},
        pages = {117},
          doi = {10.3847/1538-4357/ad2c09},
archivePrefix = {arXiv},
       eprint = {2403.02111},
 primaryClass = {astro-ph.CO},
       adsurl = {https://ui.adsabs.harvard.edu/abs/2024ApJ...965..117B}
}

@ARTICLE{Peluso2022,
       author = {{Peluso}, Giorgia and {Vulcani}, Benedetta and {Poggianti}, Bianca M. and {Moretti}, Alessia and {Radovich}, Mario and {Smith}, Rory and {Jaff{\'e}}, Yara L. and {Crossett}, Jacob and {Gullieuszik}, Marco and {Fritz}, Jacopo and {Ignesti}, Alessandro},
        title = "{Exploring the AGN-Ram Pressure Stripping Connection in Local Clusters}",
      journal = {\apj},
         year = 2022,
        month = mar,
       volume = {927},
       number = {1},
          eid = {130},
        pages = {130},
          doi = {10.3847/1538-4357/ac4225},
archivePrefix = {arXiv},
       eprint = {2111.02538},
 primaryClass = {astro-ph.GA},
       adsurl = {https://ui.adsabs.harvard.edu/abs/2022ApJ...927..130P}
}

@ARTICLE{Lourenco2023,
       author = {{Louren{\c{c}}o}, Ana C.~C. and {Jaff{\'e}}, Y.~L. and {Vulcani}, B. and {Biviano}, A. and {Poggianti}, B. and {Moretti}, A. and {Kelkar}, K. and {Crossett}, J.~P. and {Gitti}, M. and {Smith}, R. and {Lagan{\'a}}, T.~F. and {Gullieuszik}, M. and {Ignesti}, A. and {McGee}, S. and {Wolter}, A. and {Sonkamble}, S. and {M{\"u}ller}, A.},
        title = "{The effect of cluster dynamical state on ram-pressure stripping}",
      journal = {\mnras},
         year = 2023,
        month = dec,
       volume = {526},
       number = {4},
        pages = {4831-4847},
          doi = {10.1093/mnras/stad2972},
archivePrefix = {arXiv},
       eprint = {2309.15934},
 primaryClass = {astro-ph.GA},
       adsurl = {https://ui.adsabs.harvard.edu/abs/2023MNRAS.526.4831L}
}

@ARTICLE{Roberts2022,
       author = {{Roberts}, I.~D. and {van Weeren}, R.~J. and {Timmerman}, R. and {Botteon}, A. and {Gendron-Marsolais}, M. and {Ignesti}, A. and {Rottgering}, H.~J.~A.},
        title = "{LoTSS jellyfish galaxies. III. The first identification of jellyfish galaxies in the Perseus cluster}",
      journal = {\aap},
         year = 2022,
        month = feb,
       volume = {658},
          eid = {A44},
        pages = {A44},
          doi = {10.1051/0004-6361/202142294},
archivePrefix = {arXiv},
       eprint = {2112.08728},
 primaryClass = {astro-ph.GA},
       adsurl = {https://ui.adsabs.harvard.edu/abs/2022A&A...658A..44R}
}

@ARTICLE{Yagi2010,
       author = {{Yagi}, Masafumi and {Yoshida}, Michitoshi and {Komiyama}, Yutaka and {Kashikawa}, Nobunari and {Furusawa}, Hisanori and {Okamura}, Sadanori and {Graham}, Alister W. and {Miller}, Neal A. and {Carter}, David and {Mobasher}, Bahram and {Jogee}, Shardha},
        title = "{A Dozen New Galaxies Caught in the Act: Gas Stripping and Extended Emission Line Regions in the Coma Cluster}",
      journal = {\aj},
         year = 2010,
        month = dec,
       volume = {140},
       number = {6},
        pages = {1814-1829},
          doi = {10.1088/0004-6256/140/6/1814},
archivePrefix = {arXiv},
       eprint = {1005.3874},
 primaryClass = {astro-ph.CO},
       adsurl = {https://ui.adsabs.harvard.edu/abs/2010AJ....140.1814Y}
}

@ARTICLE{Vulcani2020b,
       author = {{Vulcani}, Benedetta and {Poggianti}, Bianca M. and {Tonnesen}, Stephanie and {McGee}, Sean L. and {Moretti}, Alessia and {Fritz}, Jacopo and {Gullieuszik}, Marco and {Jaff{\'e}}, Yara L. and {Franchetto}, Andrea and {Tomi{\v{c}}i{\'c}}, Neven and {Mingozzi}, Matilde and {Bettoni}, Daniela and {Wolter}, Anna},
        title = "{GASP XXX. The Spatially Resolved SFR-Mass Relation in Stripping Galaxies in the Local Universe}",
      journal = {\apj},
         year = 2020,
        month = aug,
       volume = {899},
       number = {2},
          eid = {98},
        pages = {98},
          doi = {10.3847/1538-4357/aba4ae},
archivePrefix = {arXiv},
       eprint = {2007.04996},
 primaryClass = {astro-ph.GA},
       adsurl = {https://ui.adsabs.harvard.edu/abs/2020ApJ...899...98V}
}

@ARTICLE{Vulcani2023,
       author = {{Vulcani}, Benedetta and {Treu}, Tommaso and {Calabr{\`o}}, Antonello and {Fritz}, Jacopo and {Poggianti}, Bianca M. and {Bergamini}, Pietro and {Bonchi}, Andrea and {Boyett}, Kristan and {Caminha}, Gabriel B. and {Castellano}, Marco and {Dressler}, Alan and {Fontana}, Adriano and {Glazebrook}, Karl and {Grillo}, Claudio and {Malkan}, Matthew A. and {Mascia}, Sara and {Mercurio}, Amata and {Merlin}, Emiliano and {Metha}, Benjamin and {Morishita}, Takahiro and {Nanayakkara}, Themiya and {Paris}, Diego and {Roberts-Borsani}, Guido and {Rosati}, Piero and {Roy}, Namrata and {Santini}, Paola and {Trenti}, Michele and {Vanzella}, Eros and {Wang}, Xin},
        title = "{Early Results from GLASS-JWST. XX. Unveiling a Population of ``Red Excess'' Galaxies in Abell2744 and in the Coeval Field}",
      journal = {\apjl},
         year = 2023,
        month = may,
       volume = {948},
       number = {2},
          eid = {L15},
        pages = {L15},
          doi = {10.3847/2041-8213/accbc4},
archivePrefix = {arXiv},
       eprint = {2303.01115},
 primaryClass = {astro-ph.GA},
       adsurl = {https://ui.adsabs.harvard.edu/abs/2023ApJ...948L..15V}
}

@ARTICLE{Vulcani2021,
       author = {{Vulcani}, Benedetta and {Poggianti}, Bianca M. and {Moretti}, Alessia and {Franchetto}, Andrea and {Bacchini}, Cecilia and {McGee}, Sean and {Jaff{\'e}}, Yara L. and {Mingozzi}, Matilde and {Werle}, Ariel and {Tomi{\v{c}}i{\'c}}, Neven and {Fritz}, Jacopo and {Bettoni}, Daniela and {Wolter}, Anna and {Gullieuszik}, Marco},
        title = "{GASP. XXXIII. The Ability of Spatially Resolved Data to Distinguish among the Different Physical Mechanisms Affecting Galaxies in Low-density Environments}",
      journal = {\apj},
         year = 2021,
        month = jun,
       volume = {914},
       number = {1},
          eid = {27},
        pages = {27},
          doi = {10.3847/1538-4357/abf655},
archivePrefix = {arXiv},
       eprint = {2104.02089},
 primaryClass = {astro-ph.GA},
       adsurl = {https://ui.adsabs.harvard.edu/abs/2021ApJ...914...27V}
}

@ARTICLE{Cameron2011,
       author = {{Cameron}, Ewan},
        title = "{On the Estimation of Confidence Intervals for Binomial Population Proportions in Astronomy: The Simplicity and Superiority of the Bayesian Approach}",
      journal = {\pasa},
         year = 2011,
        month = jun,
       volume = {28},
       number = {2},
        pages = {128-139},
          doi = {10.1071/AS10046},
archivePrefix = {arXiv},
       eprint = {1012.0566},
 primaryClass = {astro-ph.IM},
       adsurl = {https://ui.adsabs.harvard.edu/abs/2011PASA...28..128C}
}

@ARTICLE{Blanton2011,
       author = {{Blanton}, E.~L. and {Randall}, S.~W. and {Clarke}, T.~E. and {Sarazin}, C.~L. and {McNamara}, B.~R. and {Douglass}, E.~M. and {McDonald}, M.},
        title = "{A Very Deep Chandra Observation of A2052: Bubbles, Shocks, and Sloshing}",
      journal = {\apj},
         year = 2011,
        month = aug,
       volume = {737},
       number = {2},
          eid = {99},
        pages = {99},
          doi = {10.1088/0004-637X/737/2/99},
archivePrefix = {arXiv},
       eprint = {1105.4572},
 primaryClass = {astro-ph.CO},
       adsurl = {https://ui.adsabs.harvard.edu/abs/2011ApJ...737...99B}
}

@ARTICLE{Bellhouse2022,
       author = {{Bellhouse}, Callum and {Poggianti}, Bianca and {Moretti}, Alessia and {Vulcani}, Benedetta and {Werle}, Ariel and {Gullieuszik}, Marco and {Radovich}, Mario and {Jaff{\'e}}, Yara and {Fritz}, Jacopo and {Ignesti}, Alessandro and {Bacchini}, Cecilia and {Tomi{\v{c}}i{\'c}}, Neven and {Richard}, Johan and {Soucail}, Genevi{\`e}ve},
        title = "{Locations and Morphologies of Jellyfish Galaxies in A2744 and A370}",
      journal = {\apj},
         year = 2022,
        month = sep,
       volume = {937},
       number = {1},
          eid = {18},
        pages = {18},
          doi = {10.3847/1538-4357/ac8b6e},
archivePrefix = {arXiv},
       eprint = {2208.10524},
 primaryClass = {astro-ph.GA},
       adsurl = {https://ui.adsabs.harvard.edu/abs/2022ApJ...937...18B}
}

@ARTICLE{Stroe2015,
       author = {{Stroe}, Andra and {Oosterloo}, Tom and {R{\"o}ttgering}, Huub J.~A. and {Sobral}, David and {van Weeren}, Reinout and {Dawson}, William},
        title = "{Neutral hydrogen gas, past and future star formation in galaxies in and around the `Sausage' merging galaxy cluster}",
      journal = {\mnras},
         year = 2015,
        month = sep,
       volume = {452},
       number = {3},
        pages = {2731-2744},
          doi = {10.1093/mnras/stv1462},
archivePrefix = {arXiv},
       eprint = {1506.08822},
 primaryClass = {astro-ph.GA},
       adsurl = {https://ui.adsabs.harvard.edu/abs/2015MNRAS.452.2731S}
}

@ARTICLE{Sun2006,
       author = {{Sun}, M. and {Jones}, C. and {Forman}, W. and {Nulsen}, P.~E.~J. and {Donahue}, M. and {Voit}, G.~M.},
        title = "{A 70 Kiloparsec X-Ray Tail in the Cluster A3627}",
      journal = {\apjl},
         year = 2006,
        month = feb,
       volume = {637},
       number = {2},
        pages = {L81-L84},
          doi = {10.1086/500590},
archivePrefix = {arXiv},
       eprint = {astro-ph/0511516},
 primaryClass = {astro-ph},
       adsurl = {https://ui.adsabs.harvard.edu/abs/2006ApJ...637L..81S}
}

@ARTICLE{Ignesti2023,
       author = {{Ignesti}, A. and {Vulcani}, B. and {Botteon}, A. and {Poggianti}, B. and {Giunchi}, E. and {Smith}, R. and {Brunetti}, G. and {Roberts}, I.~D. and {van Weeren}, R.~J. and {Rajpurohit}, K.},
        title = "{Radio continuum tails in ram pressure-stripped spiral galaxies: Experimenting with a semi-empirical model in Abell 2255}",
      journal = {\aap},
         year = 2023,
        month = jul,
       volume = {675},
          eid = {A118},
        pages = {A118},
          doi = {10.1051/0004-6361/202346517},
archivePrefix = {arXiv},
       eprint = {2305.19941},
 primaryClass = {astro-ph.GA},
       adsurl = {https://ui.adsabs.harvard.edu/abs/2023A&A...675A.118I}
}

@ARTICLE{Giunchi2023a,
       author = {{Giunchi}, Eric and {Gullieuszik}, Marco and {Poggianti}, Bianca M. and {Moretti}, Alessia and {Werle}, Ariel and {Scarlata}, Claudia and {Zanella}, Anita and {Vulcani}, Benedetta and {Calzetti}, Daniela},
        title = "{HST Imaging of Star-forming Clumps in Six GASP Ram-pressure-stripped Galaxies}",
      journal = {\apj},
         year = 2023,
        month = jun,
       volume = {949},
       number = {2},
          eid = {72},
        pages = {72},
          doi = {10.3847/1538-4357/acc5ee},
archivePrefix = {arXiv},
       eprint = {2302.10615},
 primaryClass = {astro-ph.GA},
       adsurl = {https://ui.adsabs.harvard.edu/abs/2023ApJ...949...72G}
}

@ARTICLE{Giunchi2023b,
       author = {{Giunchi}, Eric and {Poggianti}, Bianca M. and {Gullieuszik}, Marco and {Moretti}, Alessia and {Werle}, Ariel and {Zanella}, Anita and {Vulcani}, Benedetta and {Tonnesen}, Stephanie and {Calzetti}, Daniela and {Bellhouse}, Callum and {Scarlata}, Claudia and {Bacchini}, Cecilia},
        title = "{Morphology of Star-forming Clumps in Ram-pressure Stripped Galaxies as Seen by HST}",
      journal = {\apj},
         year = 2023,
        month = nov,
       volume = {958},
       number = {1},
          eid = {73},
        pages = {73},
          doi = {10.3847/1538-4357/acfed6},
archivePrefix = {arXiv},
       eprint = {2310.07267},
 primaryClass = {astro-ph.GA},
       adsurl = {https://ui.adsabs.harvard.edu/abs/2023ApJ...958...73G}
}

@ARTICLE{Giunchi2025,
       author = {{Giunchi}, Eric and {Scarlata}, Claudia and {Werle}, Ariel and {Poggianti}, Bianca M. and {Moretti}, Alessia and {Gullieuszik}, Marco and {Vulcani}, Benedetta and {Ignesti}, Alessandro and {Marasco}, Antonino and {Zanella}, Anita and {Wolter}, Anna},
        title = "{Clump formation in ram-pressure stripped galaxies: Evidence from mass function}",
      journal = {\aap},
         year = 2025,
        month = apr,
       volume = {696},
          eid = {A228},
        pages = {A228},
          doi = {10.1051/0004-6361/202452983},
archivePrefix = {arXiv},
       eprint = {2502.15554},
 primaryClass = {astro-ph.GA},
       adsurl = {https://ui.adsabs.harvard.edu/abs/2025A&A...696A.228G}
}

@ARTICLE{Vogelsberger2014,
       author = {{Vogelsberger}, Mark and {Genel}, Shy and {Springel}, Volker and {Torrey}, Paul and {Sijacki}, Debora and {Xu}, Dandan and {Snyder}, Greg and {Nelson}, Dylan and {Hernquist}, Lars},
        title = "{Introducing the Illustris Project: simulating the coevolution of dark and visible matter in the Universe}",
      journal = {\mnras},
         year = 2014,
        month = oct,
       volume = {444},
       number = {2},
        pages = {1518-1547},
          doi = {10.1093/mnras/stu1536},
archivePrefix = {arXiv},
       eprint = {1405.2921},
 primaryClass = {astro-ph.CO},
       adsurl = {https://ui.adsabs.harvard.edu/abs/2014MNRAS.444.1518V}
}

@ARTICLE{Genel2014,
       author = {{Genel}, Shy and {Vogelsberger}, Mark and {Springel}, Volker and {Sijacki}, Debora and {Nelson}, Dylan and {Snyder}, Greg and {Rodriguez-Gomez}, Vicente and {Torrey}, Paul and {Hernquist}, Lars},
        title = "{Introducing the Illustris project: the evolution of galaxy populations across cosmic time}",
      journal = {\mnras},
         year = 2014,
        month = nov,
       volume = {445},
       number = {1},
        pages = {175-200},
          doi = {10.1093/mnras/stu1654},
archivePrefix = {arXiv},
       eprint = {1405.3749},
 primaryClass = {astro-ph.CO},
       adsurl = {https://ui.adsabs.harvard.edu/abs/2014MNRAS.445..175G}
}

@ARTICLE{Sijacki2015,
       author = {{Sijacki}, Debora and {Vogelsberger}, Mark and {Genel}, Shy and {Springel}, Volker and {Torrey}, Paul and {Snyder}, Gregory F. and {Nelson}, Dylan and {Hernquist}, Lars},
        title = "{The Illustris simulation: the evolving population of black holes across cosmic time}",
      journal = {\mnras},
         year = 2015,
        month = sep,
       volume = {452},
       number = {1},
        pages = {575-596},
          doi = {10.1093/mnras/stv1340},
archivePrefix = {arXiv},
       eprint = {1408.6842},
 primaryClass = {astro-ph.GA},
       adsurl = {https://ui.adsabs.harvard.edu/abs/2015MNRAS.452..575S}
}

@ARTICLE{Raddick2013,
       author = {{Raddick}, M. Jordan and {Bracey}, Georgia and {Gay}, Pamela L. and {Lintott}, Chris J. and {Cardamone}, Carie and {Murray}, Phil and {Schawinski}, Kevin and {Szalay}, Alexander S. and {Vandenberg}, Jan},
        title = "{Galaxy Zoo: Motivations of Citizen Scientists}",
      journal = {arXiv e-prints},
         year = 2013,
        month = mar,
          eid = {arXiv:1303.6886},
        pages = {arXiv:1303.6886},
          doi = {10.48550/arXiv.1303.6886},
archivePrefix = {arXiv},
       eprint = {1303.6886},
 primaryClass = {physics.ed-ph},
       adsurl = {https://ui.adsabs.harvard.edu/abs/2013arXiv1303.6886R}
}

@ARTICLE{Sampaio2024,
       author = {{Sampaio}, V.~M. and {de Carvalho}, R.~R. and {Arag{\'o}n-Salamanca}, A. and {Merrifield}, M.~R. and {Ferreras}, I. and {Cornwell}, D.~J.},
        title = "{Exploring galaxy evolution time-scales in clusters: insights from the projected phase space}",
      journal = {\mnras},
         year = 2024,
        month = jul,
       volume = {532},
       number = {1},
        pages = {982-994},
          doi = {10.1093/mnras/stae1533},
archivePrefix = {arXiv},
       eprint = {2406.12273},
 primaryClass = {astro-ph.GA},
       adsurl = {https://ui.adsabs.harvard.edu/abs/2024MNRAS.532..982S}
}

@ARTICLE{George2024,
       author = {{George}, K. and {Poggianti}, B.~M. and {Omizzolo}, A. and {Vulcani}, B. and {C{\^o}t{\'e}}, P. and {Postma}, J. and {Smith}, R. and {Jaffe}, Y.~L. and {Gullieuszik}, M. and {Moretti}, A. and {Subramaniam}, A. and {Sreekumar}, P. and {Ghosh}, S.~K. and {Tandon}, S.~N. and {Hutchings}, J.~B.},
        title = "{Candidate ram-pressure stripped galaxies in six low-redshift clusters revealed from ultraviolet imaging}",
      journal = {\aap},
         year = 2024,
        month = oct,
       volume = {690},
          eid = {A337},
        pages = {A337},
          doi = {10.1051/0004-6361/202450302},
archivePrefix = {arXiv},
       eprint = {2409.10586},
 primaryClass = {astro-ph.GA},
       adsurl = {https://ui.adsabs.harvard.edu/abs/2024A&A...690A.337G}
}

@ARTICLE{Matijevic2026,
       author = {{Matijevi{\'c}}, L. and {Tomi{\v{c}}i{\'c}}, N. and {Marasco}, A. and {Ignesti}, A. and {Lassen}, A.~E. and {Smith}, R. and {Sell}, P. and {Roberts}, I.~D. and {Zezas}, A. and {Anastasopoulou}, K. and {Kotoulas}, P. and {Ba{\v{s}}i{\'c}}, R.},
        title = "{The competing influences of ram pressure and tidal interaction in NGC 2276}",
      journal = {\aap},
         year = 2026,
        month = feb,
       volume = {707},
          eid = {A40},
        pages = {A40},
          doi = {10.1051/0004-6361/202555393},
archivePrefix = {arXiv},
       eprint = {2512.17486},
 primaryClass = {astro-ph.GA},
       adsurl = {https://ui.adsabs.harvard.edu/abs/2026A&A...707A..40M}
}

\begin{appendix}

\twocolumn
\section{Redshift of the cluster sample}

Figure~\ref{fig:z_dist} shows the distribution of redshifts of the clusters considered in our sample, which is representative of clusters across the halo mass range considered, with a slight undersampling at very low redshifts ($z < 0.025$).

\begin{figure}[ht!]
    \centering
    \includegraphics[width=0.49\textwidth]{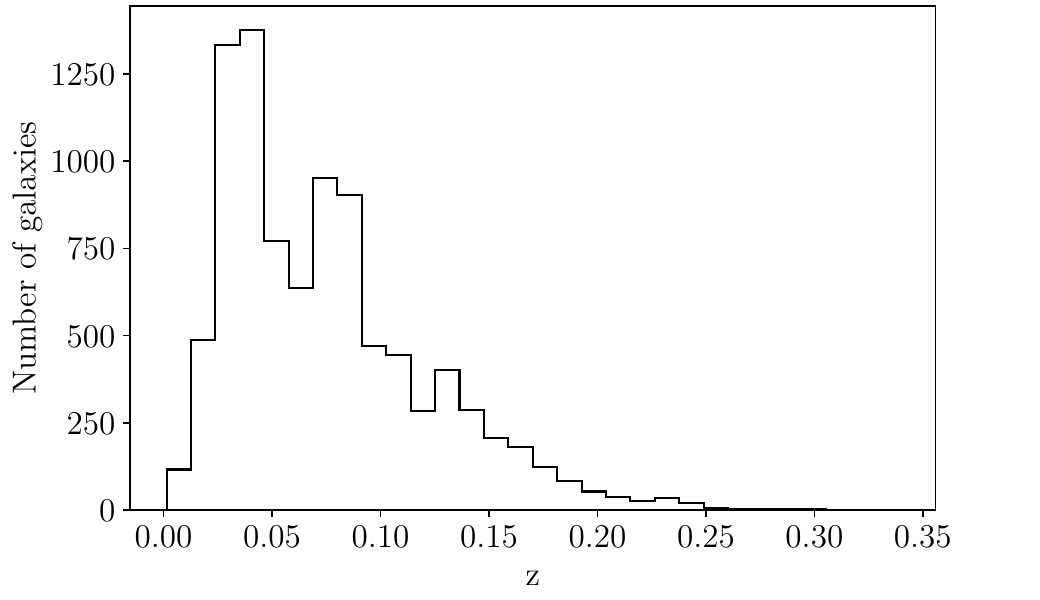}
    \caption{Redshift distribution of the clusters listed in Tables~\ref{tab:dr9_clusters} and \ref{tab:dr10_clusters}}
    \label{fig:z_dist}
\end{figure}

\section{Number of classifications per galaxy}

Figure~\ref{fig:classifications_hist} shows the distribution of classifications per galaxy, showing the minimum number of classifications is 10, the maximum is above 70, and that the sample is not dominated by poorly sampled galaxies. 

\begin{figure}[h!]
    \centering
    \includegraphics[width=0.49\textwidth]{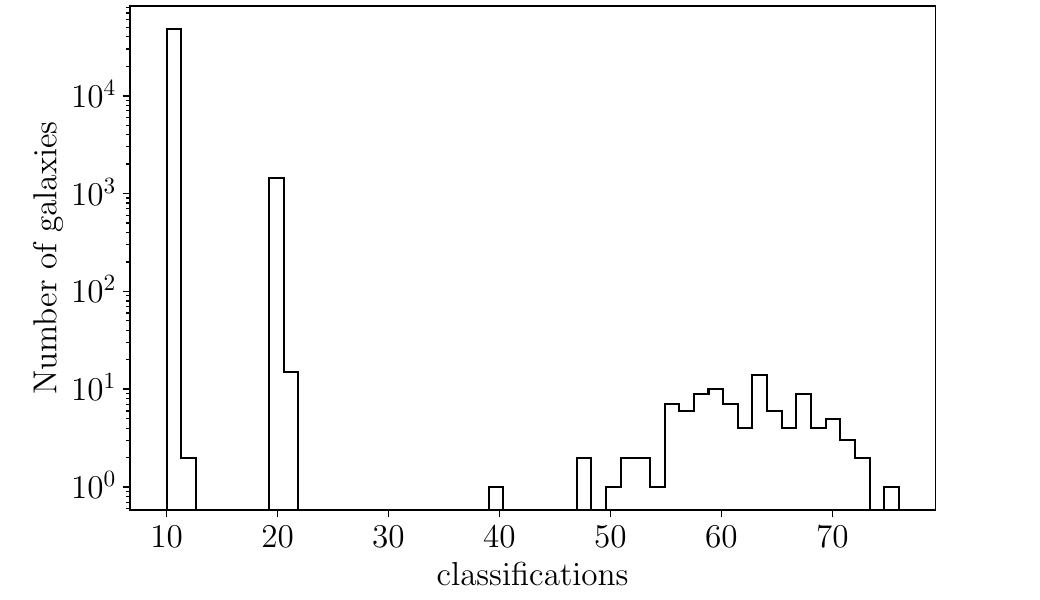}
    \caption{Distribution of classifications per galaxy, showing that the sample is not dominated by poorly sampled galaxies.  
}
    \label{fig:classifications_hist}
\end{figure}

\onecolumn
\section{Classification workflow steps}
\begin{figure*}[ht!]
    \centering
    \includegraphics[width=0.8\textwidth]{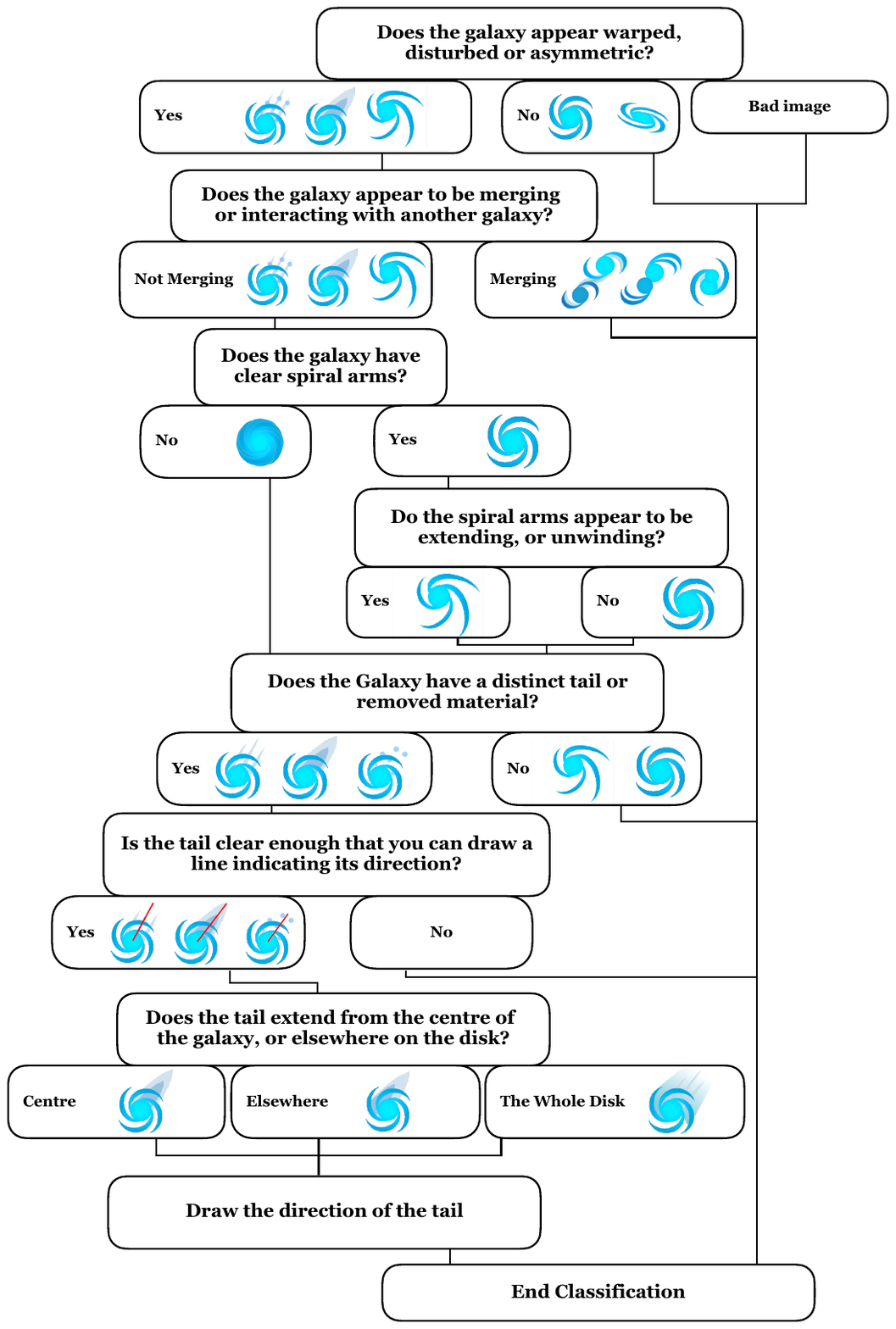}
    \caption{Workflow used in the classification process. The workflow is designed to minimise the time spent classifying objects, such as undisturbed galaxies or mergers, which are not of interest, whilst simultaneously maximising the information gained from objects of interest, such as those with signs of stripped tails.}
    \label{fig:workflow_diagram}
\end{figure*}

\section{Classification criteria optimisation}
\begin{figure*}[ht!]
    \centering
    \includegraphics[width=\linewidth]{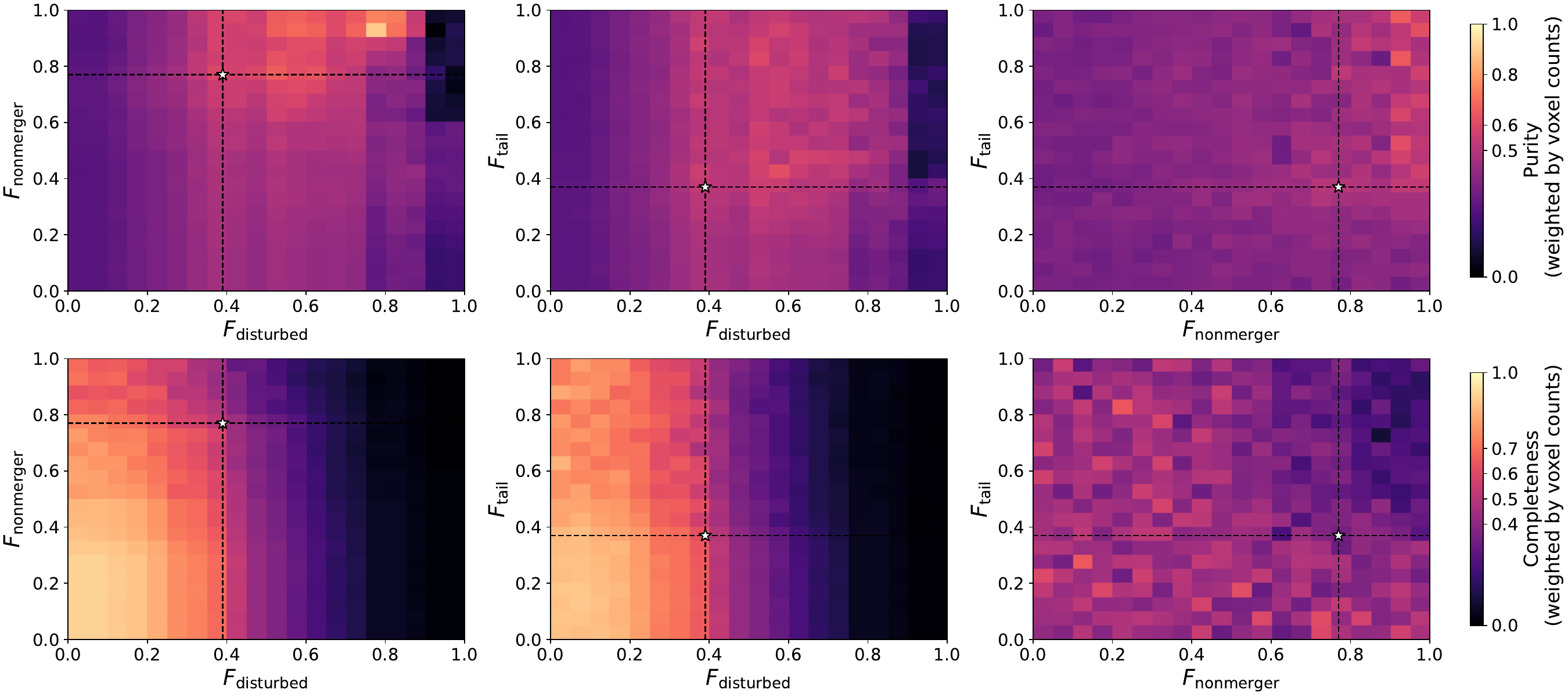}
    \caption{Distribution of purity (top) and completeness (bottom) for different combinations of the key vote fractions used for selecting RPS candidates: $(F_{\mathrm{disturbed}},\,F_{\mathrm{non\_merger}},$ and$\,F_{\mathrm{tail}})$.
    The plots are voxelised with uniform cells $\Delta x=\Delta y=\Delta z=0.05$. Within each voxel we compute the mean of the metric, and each 2D pixel displays in colour the count-weighted mean across the hidden axis. The dashed lines mark the adopted thresholds in our criteria $(0.39,\,0.77,\,0.37)$ (see Sect.~\ref{sec:criteria}), and the star marks their intersection. }
    \label{fig:Monte_carlo_plot}
\end{figure*}

\section{Catalogue description}

Table~\ref{table:catalogue} shows a brief example of the catalogue of classified objects associated with this paper. The selection criteria to produce the samples of SC, SC+T, and MC galaxies in this study are described in Sect.\,\ref{sec:criteria}; however, for most purposes we recommend recalculating appropriate criteria depending on the level of purity required.

\begin{landscape}
\begin{table}\small
    \centering
    \caption{Example lines from the catalogue showing a selection of classified galaxies from two separate regions.}
    \begin{tabular}{S[round-mode=places,round-precision=3] S[round-mode=places,round-precision=3] S[round-mode=places,round-precision=0,table-format=2] F F F F F F F F W V W V W V W V}
        {RA} & {Dec} & {$\mathrm{N}_\mathrm{class.}$} & \disturbedfrac{} & \disturbedfrac{} & \mergerfrac{} & \mergerfrac{} & \tailfrac{} & \tailfrac{} & \problemfrac{} & \problemfrac{} & \disturbedvotes{} & \disturbedvotes{} & \mergervotes{} & \mergervotes{} & \tailvotes{} & \tailvotes{} & \problemvotes{} & \problemvotes{} \\ 
         & & & {\textit{\scriptsize weighted}} && {\textit{\scriptsize weighted}} & & {\textit{\scriptsize weighted}} & & {\textit{\scriptsize weighted}} &  & {\textit{\scriptsize weighted}} & & {\textit{\scriptsize weighted}} & & {\scriptsize weighted} & & {\textit{\scriptsize weighted}} &\\ \hline
        90.44867716946699 &	-59.37549536319105 &	10.0 &	0.125 &	0.125 &	0.0 &	0.0 &	1.0 &	1.0 &	0.2 &	0.2 &	1.0 &	1.0 &	0.0 &	0.0 &	1.0 &	1.0 &	2.0 &	2.0 \\
89.29648582890579 &	-59.20671280594229 &	10.0 &	0.1 &	0.1 &	0.0 &	0.0 &	1.0 &	1.0 &	0.0 &	0.0 &	1.0 &	1.0 &	0.0 &	0.0 &	1.0 &	1.0 &	0.0 &	0.0 \\
89.39255001585155 &	-59.24499324559322 &	10.0 &	0.0 &	0.0 &	0.0 &	0.0 &	0.0 &	0.0 &	0.3 &	0.3 &	0.0 &	0.0 &	0.0 &	0.0 &	0.0 &	0.0 &	3.0 &	3.0 \\
89.62162305315748 &	-59.315210387981345 &	10.0 &	0.1 &	0.1 &	0.0 &	0.0 &	1.0 &	1.0 &	0.0 &	0.0 &	1.0 &	1.0 &	0.0 &	0.0 &	1.0 &	1.0 &	0.0 &	0.0 \\
89.62260781221386 &	-59.12508636380114 &	10.0 &	0.0 &	0.0 &	0.0 &	0.0 &	0.0 &	0.0 &	0.0 &	0.0 &	0.0 &	0.0 &	0.0 &	0.0 &	0.0 &	0.0 &	0.0 &	0.0 \\
89.68647164888881 &	-59.22125569476057 &	10.0 &	0.16666666666666666 &	0.2 &	0.0 &	0.0 &	0.0 &	0.0 &	0.0 &	0.0 &	1.5 &	2.0 &	0.0 &	0.0 &	0.0 &	0.0 &	0.0 &	0.0 \\
89.74224476183839 &	-59.19056660760814 &	10.0 &	0.42857142857142855 &	0.42857142857142855 &	0.6666666666666666 &	0.6666666666666666 &	0.0 &	0.0 &	0.3 &	0.3 &	3.0 &	3.0 &	2.0 &	2.0 &	0.0 &	0.0 &	3.0 &	3.0 \\
89.87805417138182 &	-59.3046484048759 &	10.0 &	0.25 &	0.25 &	0.0 &	0.0 &	0.5 &	0.5 &	0.2 &	0.2 &	2.0 &	2.0 &	0.0 &	0.0 &	1.0 &	1.0 &	2.0 &	2.0 \\
89.92794818495409 &	-59.19764160248818 &	10.0 &	0.375 &	0.375 &	0.3333333333333333 &	0.3333333333333333 &	0.3333333333333333 &	0.3333333333333333 &	0.2 &	0.2 &	3.0 &	3.0 &	1.0 &	1.0 &	1.0 &	1.0 &	2.0 &	2.0 \\
90.08351576253136 &	-59.36622845863936 &	10.0 &	0.0 &	0.0 &	0.0 &	0.0 &	0.0 &	0.0 &	0.2 &	0.2 &	0.0 &	0.0 &	0.0 &	0.0 &	0.0 &	0.0 &	2.0 &	2.0 \\
90.14547960412588 &	-59.235135370199735 &	10.0 &	0.3333333333333333 &	0.3333333333333333 &	0.3333333333333333 &	0.3333333333333333 &	0.6666666666666666 &	0.6666666666666666 &	0.1 &	0.1 &	3.0 &	3.0 &	1.0 &	1.0 &	2.0 &	2.0 &	1.0 &	1.0 \\
90.15164311568115 &	-59.255667239670906 &	10.0 &	0.42857142857142855 &	0.42857142857142855 &	0.6666666666666666 &	0.6666666666666666 &	0.3333333333333333 &	0.3333333333333333 &	0.3 &	0.3 &	3.0 &	3.0 &	2.0 &	2.0 &	1.0 &	1.0 &	3.0 &	3.0 \\
90.16217408551607 &	-59.14769905095829 &	10.0 &	0.3 &	0.3 &	0.0 &	0.0 &	0.6666666666666666 &	0.6666666666666666 &	0.0 &	0.0 &	3.0 &	3.0 &	0.0 &	0.0 &	2.0 &	2.0 &	0.0 &	0.0 \\
90.32180585026609 &	-59.237374793137704 &	10.0 &	0.3 &	0.3 &	0.3333333333333333 &	0.3333333333333333 &	0.6666666666666666 &	0.6666666666666666 &	0.0 &	0.0 &	3.0 &	3.0 &	1.0 &	1.0 &	2.0 &	2.0 &	0.0 &	0.0 \\
90.34048221541471 &	-59.316911511271066 &	10.0 &	0.1111111111111111 &	0.1111111111111111 &	0.0 &	0.0 &	0.0 &	0.0 &	0.1 &	0.1 &	1.0 &	1.0 &	0.0 &	0.0 &	0.0 &	0.0 &	1.0 &	1.0 \\
90.55645228782733 &	-59.357888444346244 &	10.0 &	0.5 &	0.5 &	0.0 &	0.0 &	0.8 &	0.8 &	0.0 &	0.0 &	5.0 &	5.0 &	0.0 &	0.0 &	4.0 &	4.0 &	0.0 &	0.0 \\
90.62714448965755 &	-59.15859169605256 &	10.0 &	0.0 &	0.0 &	0.0 &	0.0 &	0.0 &	0.0 &	0.2631578947368421 &	0.3 &	0.0 &	0.0 &	0.0 &	0.0 &	0.0 &	0.0 &	2.5 &	3.0 \\
90.98446546496308 &	-59.149364240590906 &	10.0 &	0.375 &	0.375 &	0.3333333333333333 &	0.3333333333333333 &	0.6666666666666666 &	0.6666666666666666 &	0.2 &	0.2 &	3.0 &	3.0 &	1.0 &	1.0 &	2.0 &	2.0 &	2.0 &	2.0 \\
91.14688505258185 &	-59.18720671971016 &	10.0 &	0.7894736842105263 &	0.8 &	0.0 &	0.0 &	0.5333333333333333 &	0.5 &	0.0 &	0.0 &	7.5 &	8.0 &	0.0 &	0.0 &	4.0 &	4.0 &	0.0 &	0.0 \\
91.19720698796995 &	-59.19625748218365 &	10.0 &	0.7 &	0.7 &	0.0 &	0.0 &	1.0 &	1.0 &	0.0 &	0.0 &	7.0 &	7.0 &	0.0 &	0.0 &	7.0 &	7.0 &	0.0 &	0.0 \\
89.41988497305825 &	-58.982283159337214 &	11.0 &	0.2727272727272727 &	0.2727272727272727 &	0.0 &	0.0 &	0.3333333333333333 &	0.3333333333333333 &	0.0 &	0.0 &	3.0 &	3.0 &	0.0 &	0.0 &	1.0 &	1.0 &	0.0 &	0.0 \\
89.49084043584381 &	-58.97999775308319 &	10.0 &	0.5555555555555556 &	0.5555555555555556 &	0.0 &	0.0 &	0.4 &	0.4 &	0.1 &	0.1 &	5.0 &	5.0 &	0.0 &	0.0 &	2.0 &	2.0 &	1.0 &	1.0 \\
89.50348036706882 &	-59.049534673976545 &	10.0 &	0.2 &	0.2 &	0.5 &	0.5 &	0.0 &	0.0 &	0.0 &	0.0 &	2.0 &	2.0 &	1.0 &	1.0 &	0.0 &	0.0 &	0.0 &	0.0 \\
89.57599452214707 &	-59.03227859551985 &	10.0 &	0.0 &	0.0 &	0.0 &	0.0 &	0.0 &	0.0 &	0.3 &	0.3 &	0.0 &	0.0 &	0.0 &	0.0 &	0.0 &	0.0 &	3.0 &	3.0 \\
89.78093593807772 &	-59.06201621219484 &	10.0 &	0.14285714285714285 &	0.14285714285714285 &	0.0 &	0.0 &	0.0 &	0.0 &	0.3 &	0.3 &	1.0 &	1.0 &	0.0 &	0.0 &	0.0 &	0.0 &	3.0 &	3.0 \\
89.87542745741641 &	-59.02159737099034 &	10.0 &	0.1 &	0.1 &	0.0 &	0.0 &	1.0 &	1.0 &	0.0 &	0.0 &	1.0 &	1.0 &	0.0 &	0.0 &	1.0 &	1.0 &	0.0 &	0.0 \\
89.95467831994915 &	-58.935373508291924 &	10.0 &	0.25 &	0.25 &	0.5 &	0.5 &	0.5 &	0.5 &	0.2 &	0.2 &	2.0 &	2.0 &	1.0 &	1.0 &	1.0 &	1.0 &	2.0 &	2.0 \\
90.60839459785471 &	-59.105373859506685 &	10.0 &	0.5 &	0.5 &	0.0 &	0.0 &	0.5 &	0.5 &	0.2 &	0.2 &	4.0 &	4.0 &	0.0 &	0.0 &	2.0 &	2.0 &	2.0 &	2.0 \\
90.73402416525542 &	-59.04003111473765 &	10.0 &	0.6666666666666666 &	0.6666666666666666 &	0.3333333333333333 &	0.3333333333333333 &	0.6666666666666666 &	0.6666666666666666 &	0.1 &	0.1 &	6.0 &	6.0 &	2.0 &	2.0 &	4.0 &	4.0 &	1.0 &	1.0 \\
353.8176232921158 &	26.317329876117817 &	10.0 &	0.3333333333333333 &	0.3333333333333333 &	0.5 &	0.5 &	0.5 &	0.5 &	0.3684210526315789 &	0.4 &	2.0 &	2.0 &	1.0 &	1.0 &	1.0 &	1.0 &	3.5 &	4.0 \\
353.82529168398315 &	26.295779229516416 &	10.0 &	0.375 &	0.375 &	0.0 &	0.0 &	0.6666666666666666 &	0.6666666666666666 &	0.15789473684210525 &	0.2 &	3.0 &	3.0 &	0.0 &	0.0 &	2.0 &	2.0 &	1.5 &	2.0 \\
353.84562221739935 &	26.147041282051703 &	10.0 &	0.4 &	0.4 &	0.5 &	0.5 &	0.0 &	0.0 &	0.5 &	0.5 &	2.0 &	2.0 &	1.0 &	1.0 &	0.0 &	0.0 &	5.0 &	5.0 \\
353.8513162859483 &	26.18154738696136 &	10.0 &	0.11764705882352941 &	0.1111111111111111 &	0.0 &	0.0 &	1.0 &	1.0 &	0.10526315789473684 &	0.1 &	1.0 &	1.0 &	0.0 &	0.0 &	1.0 &	1.0 &	1.0 &	1.0 \\
353.87404604966315 &	26.18418142779689 &	10.0 &	0.375 &	0.375 &	0.0 &	0.0 &	1.0 &	1.0 &	0.2 &	0.2 &	3.0 &	3.0 &	0.0 &	0.0 &	3.0 &	3.0 &	2.0 &	2.0 \\
353.88364082904883 &	26.314872920788808 &	10.0 &	0.2222222222222222 &	0.2222222222222222 &	0.0 &	0.0 &	1.0 &	1.0 &	0.1 &	0.1 &	2.0 &	2.0 &	0.0 &	0.0 &	2.0 &	2.0 &	1.0 &	1.0 \\
353.89724860717195 &	26.200882992021423 &	10.0 &	0.11764705882352941 &	0.1111111111111111 &	1.0 &	1.0 &	0.0 &	0.0 &	0.10526315789473684 &	0.1 &	1.0 &	1.0 &	1.0 &	1.0 &	0.0 &	0.0 &	1.0 &	1.0 \\
353.9104440666401 &	26.36709341655761 &	10.0 &	0.14285714285714285 &	0.14285714285714285 &	0.0 &	0.0 &	1.0 &	1.0 &	0.3 &	0.3 &	1.0 &	1.0 &	0.0 &	0.0 &	1.0 &	1.0 &	3.0 &	3.0 \\
354.01656821275816 &	26.36738516178934 &	10.0 &	0.0 &	0.0 &	0.0 &	0.0 &	0.0 &	0.0 &	0.6 &	0.6 &	0.0 &	0.0 &	0.0 &	0.0 &	0.0 &	0.0 &	6.0 &	6.0 \\
354.05537752261546 &	26.36188472321961 &	10.0 &	0.875 &	0.875 &	0.14285714285714285 &	0.14285714285714285 &	0.2857142857142857 &	0.2857142857142857 &	0.2 &	0.2 &	7.0 &	7.0 &	1.0 &	1.0 &	2.0 &	2.0 &	2.0 &	2.0 \\
354.0784126092739 &	26.18022564943489 &	10.0 &	0.14285714285714285 &	0.14285714285714285 &	0.0 &	0.0 &	0.0 &	0.0 &	0.3 &	0.3 &	1.0 &	1.0 &	0.0 &	0.0 &	0.0 &	0.0 &	3.0 &	3.0 \\
354.0996562221021 &	26.36991673805481 &	10.0 &	0.3 &	0.3 &	0.0 &	0.0 &	0.3333333333333333 &	0.3333333333333333 &	0.0 &	0.0 &	3.0 &	3.0 &	0.0 &	0.0 &	1.0 &	1.0 &	0.0 &	0.0 \\
$\cdots$ &$\cdots$ &$\cdots$ &$\cdots$ &$\cdots$ &$\cdots$ &$\cdots$ &$\cdots$ &$\cdots$ &$\cdots$ &$\cdots$ &$\cdots$ &$\cdots$ &$\cdots$ &$\cdots$ &$\cdots$ &$\cdots$ &$\cdots$ &$\cdots$

    \end{tabular}
    \tablefoot{$\mathrm{N}_\mathrm{class}$ is the total number of classifications, which is always greater than 10. The fractions, for example \disturbedfrac{}, are defined in Sect.\,\ref{sec:fractions}. The vote columns show the total numberfor each feature. Columns marked as \textit{weighted} show the counts after debiasing, as described in Sect.\,\ref{sec:combining}.}\label{table:catalogue}
\end{table}
\end{landscape}
\twocolumn
\label{lastpage}

\end{appendix}

\end{document}